\documentclass[aps, preprint, prc, amsmath, amssymb, nofootinbib, superscriptaddress]{revtex4}

\usepackage[usenames]{color}
\usepackage[dvipsnames]{xcolor}
\usepackage{graphicx}
\usepackage{amsmath}
\usepackage{amsfonts}
\usepackage{amssymb}
\usepackage{mathrsfs}
\usepackage{bm}
\usepackage{verbatim}
\usepackage{cancel}
\usepackage{comment}
\usepackage[normalem]{ulem}
\usepackage{CJK}
\usepackage{float}
\usepackage{textcomp}

\usepackage[utf8]{inputenc}
\usepackage{subfigure}
\usepackage{hyperref}

\hypersetup{
  colorlinks=true,        
   linkcolor=ForestGreen,    
}

\newcommand{\mhi}{M_{\text{hi}}}

\newcommand{\chn}[3]{{{}^{#1}\!{#2}_{#3}}}
\newcommand{\cs}[2]{\chn{#1}{S}{#2}}

\newcommand{\cd}[2]{\chn{#1}{D}{#2}}

\newcommand{{\NNLO}}{N$^2$LO}
\newcommand{\NNNLO}{N$^3$LO}

\graphicspath{{./figs/}}

\begin{document}

\begin{CJK*}{UTF8}{gbsn}

\title{Perturbative calculations of deuteron form factors}
\author{Wenchao Shi (史文超)}
\affiliation{College of Physics, Sichuan University, Chengdu, Sichuan 610065, China}

\author{Rui Peng (彭锐)}
\affiliation{College of Physics, Sichuan University, Chengdu, Sichuan 610065, China}

\author{Tai-Xing Liu (刘太兴)}
\affiliation{College of Physics, Sichuan University, Chengdu, Sichuan 610065, China}

\author{Songlin Lyu (吕松林)}
\email{songlin@scu.edu.cn}
\affiliation{College of Physics, Sichuan University, Chengdu, Sichuan 610065, China}

\author{Bingwei Long (龙炳蔚)}
\email{bingwei@scu.edu.cn}
\affiliation{College of Physics, Sichuan University, Chengdu, Sichuan 610065, China}

\date{July 3, 2022}

\begin{abstract}
We calculate the deuteron charge, electric quadrupole, and magnetic form factors up to next-to-next-to-leading order in chiral effective field theory, treating subleading corrections, especially that of chiral nuclear forces, in perturbation theory. We examine the power counting based on naive dimensional analysis by investigating the ultraviolet cutoff variation of these form factors. We find that the {\NNLO} magnetic form factor shows significant cutoff dependence, suggesting the contact current operator responsible be enhanced. After promotion to {\NNLO}, it indeed renormalizes the magnetic form factor. This is in agreement with a previous work based on renormalization-group analysis. For the charge and quadrupole moments, perturbative calculation allows us to study how they scale with multiple low-energy parameters such as the pion mass, deuteron binding momentum, and momentum transfer.
\end{abstract}

\maketitle

\end{CJK*}

\section{Introduction\label{sec:intro}}

Power counting of chiral effective field theory (ChEFT) in nuclear physics can be at times difficult to grasp because naive dimensional analysis (NDA)~\cite{Weinberg:1991um}, once a reliable guidance for power counting in pionic dynamics and the single-nucleon sector~\cite{Weinberg:1978kz, Gasser:1983yg, Gasser:1987rb, Bernard:1995dp}, fails to satisfy renormalization group (RG) invariance~\cite{Kaplan:1998we, Nogga:2005fv, Birse:2005um, Birse:2009my, Valderrama:2009ei, Long:2011xw, Long:2011qx, PavonValderrama:2011fcz} or to accommodate emergence of unexpected large length scales~\cite{Steele:1998zc, Long:2013cya, SanchezSanchez:2017tws, Peng:2021pvo}. In contrast with the amount of works on analyzing strong interactions, there have been only a few critical inspections of NDA-based counting of nuclear electroweak currents~\cite{PavonValderrama:2014zeq, Cirigliano:2018hja, Oosterhof:2019dlo}. In the present paper, we study power counting of electromagnetic charge and current operators in ChEFT by investigating the ultraviolet (UV) cutoff dependence of the deuteron charge ($G_C$), electric quadrupole ($G_Q$), and magnetic ($G_M$) form factors while treating subleading nuclear potentials in perturbation theory.

Because charge and current operators are defined by perturbative Feynman diagrams just like in the single-nucleon sector, one can use NDA to count powers of momenta, generically denoted by $P$, that float around in the diagram and the pion mass $m_\pi$. Derivation of nuclear electroweak charge and current operators in the framework of ChEFT has reached quite high orders by NDA counting~\cite{Park:1995pn, Pastore:2008ui, Pastore:2009is, Kolling:2009iq, Kolling:2011mt, Pastore:2011ip, Krebs:2019aka}. When inserted between initial and final states, however, low-energy constants (LECs) of these operators could be renormalized by nonperturbative nuclear interactions so significantly that NDA may not be applicable. Reference~\cite{PavonValderrama:2014zeq} pioneered RG analyses on charge and current operators, relying on the short-distance behavior of nonperturbative $NN$ wave functions generated by leading-order (LO) chiral forces. The conclusion includes a list of contact current operators for which power counting needs to be modified away from NDA. 

We will focus on the electromagnetic currents, using the deuteron charge, electric quadrupole, and magnetic form factors as the observables to carry out the RG analysis. Our methodology starts with the proposition that NDA is appropriate for power counting long-range contributions, which up to next-to-next-to-leading order ({\NNLO}) turn out to consist of mostly one-body currents. The cutoff independence of, or the lack thereof, these contributions are then examined.

Previous ChEFT calculations of the deuteron form factors were performed with NDA-based potentials and currents~\cite{Phillips:1999am, Walzl:2001vb, Phillips:2003jz, Phillips:2006im, Valderrama:2007ja, Epelbaum:2013naa, Filin:2020tcs}. The deuteron wave function is the only nuclear-force input for these form factors, so only the $\cs{3}{1}\text{-}\cd{3}{1}$ chiral potentials will be invoked. Power counting for $\cs{3}{1}\text{-}\cd{3}{1}$ is not modified by the guideline of RG invariance, at least not up to {\NNLO}~\cite{Valderrama:2009ei, Long:2011xw}; therefore, our emphasis on perturbative treatment of subleading potentials is the principle difference between previous calculations and ours. Very much like textbook examples in quantum field theory, higher-order diagrams are not necessarily small before renormalized properly, and treating higher-order corrections in perturbation theory on top of nonperturbative LO helps identify the counterterms that can renormalize them. In fact, RG analyses on subleading chiral potentials were mostly carried out in perturbation theory~\cite{Valderrama:2009ei, Long:2011xw, Long:2011qx, PavonValderrama:2011fcz, vanKolck:2020llt, Long:2012ve, Song:2016ale, PavonValderrama:2019lsu, Yang:2020pgi, Yang:2021vxa}. In addition, although only valid for smaller momenta, perturbative treatment of subleading corrections was clearly displayed in pionless and perturbative-pion calculations of the deuteron form factors~\cite{Chen:1999tn, Chen:1999vd, Kaplan:1998sz, Savage:1999cm, deVries:2011re, Mereghetti:2013bta}.

In Sec.~\ref{sec:pot-dwf} we review the chiral potentials and the deuteron wave function to be used in the form-factor calculations. In Sec.~\ref{sec:formalism} the relevant charge and current operators, and the integrals for the desired matrix elements are derived. We show and discuss the results in Sec.~\ref{sec:results}, followed by a summary in Sec.~\ref{sec:summary}.

\section{The deuteron wave function\label{sec:pot-dwf}}

We follow the power counting proposed in Ref.~\cite{Long:2011xw} to assemble the chiral forces in the coupled channel of $\cs{3}{1}\text{-}\cd{3}{1}$, which are responsible for constructing the deuteron wave function. The LO potential is made up of one-pion exchange (OPE) and the contact part. OPE is given by
\begin{equation}
    V_{1\pi}(\vec{p}\,', \vec{p}\,) = - \frac{g_A^2}{4f_\pi^2}\, \bm{\tau}_1 \bm{\cdot} \bm{\tau}_2\, \frac{\vec{\sigma}_1 \cdot (\vec{p}\,' - \vec{p}\,)\, \vec{\sigma}_2 \cdot (\vec{p}\,' - \vec{p}\,)}{(\vec{p}\,' - \vec{p}\,)^2 + m_\pi^2}  \, ,
\label{eqn:OPEpot}
\end{equation}
where $\vec{p}$ ($\vec{p}\,'$) is the incoming (outgoing) relative momentum, $m_\pi = 138.0$ MeV the pion mass, $f_\pi = 92.4$ MeV the pion decay constant, and $g_A = 1.29$ the axial-vector coupling constant. The contact potential will be denoted by a $2 \times 2$ matrix acting on $\cs{3}{1}$ and $\cd{3}{1}$:
\begin{equation}
    \left\langle \cs{3}{1}\text{-}\cd{3}{1} \left| V_S^{(0)} \right| \cs{3}{1}\text{-}\cd{3}{1} \right\rangle =
    \begin{pmatrix}
     C^{(0)} & 0 \\
     0 & 0
    \end{pmatrix} \, .
\label{eqn:LOVS}
\end{equation}

The next-to-leading order (NLO) potential vanishes and the {\NNLO} potential includes the leading deltaless, two-pion exchange (TPE0) and contact terms with two powers of momenta. The expression for TPE0 potential can be found in Ref.~\cite{Kaiser:1997mw}, and the {\NNLO} contact term has the following form~\cite{Long:2011xw}:
\begin{equation}
    \left\langle \cs{3}{1}\text{-}\cd{3}{1} \left| V_S^{(2)} \right| \cs{3}{1}\text{-}\cd{3}{1} \right\rangle =
    \begin{pmatrix}
     C^{(2)} + D^{(0)} \left( p^{\prime  2} + p^2 \right) & -E^{(0)}\, p^2 \\
     -E^{(0)}\, p^{\prime  2} & 0
    \end{pmatrix} \, .
    \label{eqn:VSN2LO}
\end{equation}
The UV part of the potentials will be regularized by a separable Gaussian regulator:
\begin{equation}
  V_{\Lambda}(\vec{p}\,', \vec{p}\,) \equiv \exp\left(-\frac{\vec{p}\,^{\prime  4}}{\Lambda^4}\right)
  V(\vec{p}\,',\vec{p}\,) \exp\left(-\frac{\vec{p}\,^4}{\Lambda^4}\right) \, ,
\end{equation}
where $\Lambda$ is the ultraviolet momentum cutoff.

The values of these contact couplings are determined as follows. At LO, $C^{(0)}$ is adjusted in order to reproduce the deuteron binding energy $B_d = 2.225$ MeV. At {\NNLO}, in addition to $B_d$ the Nijmegen empirical values~\cite{Stoks:1993tb} for the $\cs{3}{1}$ phase shift $\delta_\cs{3}{1}$ at the center-of-mass (CM) momentum $k = 120.64$ MeV and the mixing angle $\varepsilon_1$ at $k = 150.12$ MeV are fitted too.

\begin{figure}
    \centering
    \includegraphics[scale=0.4]{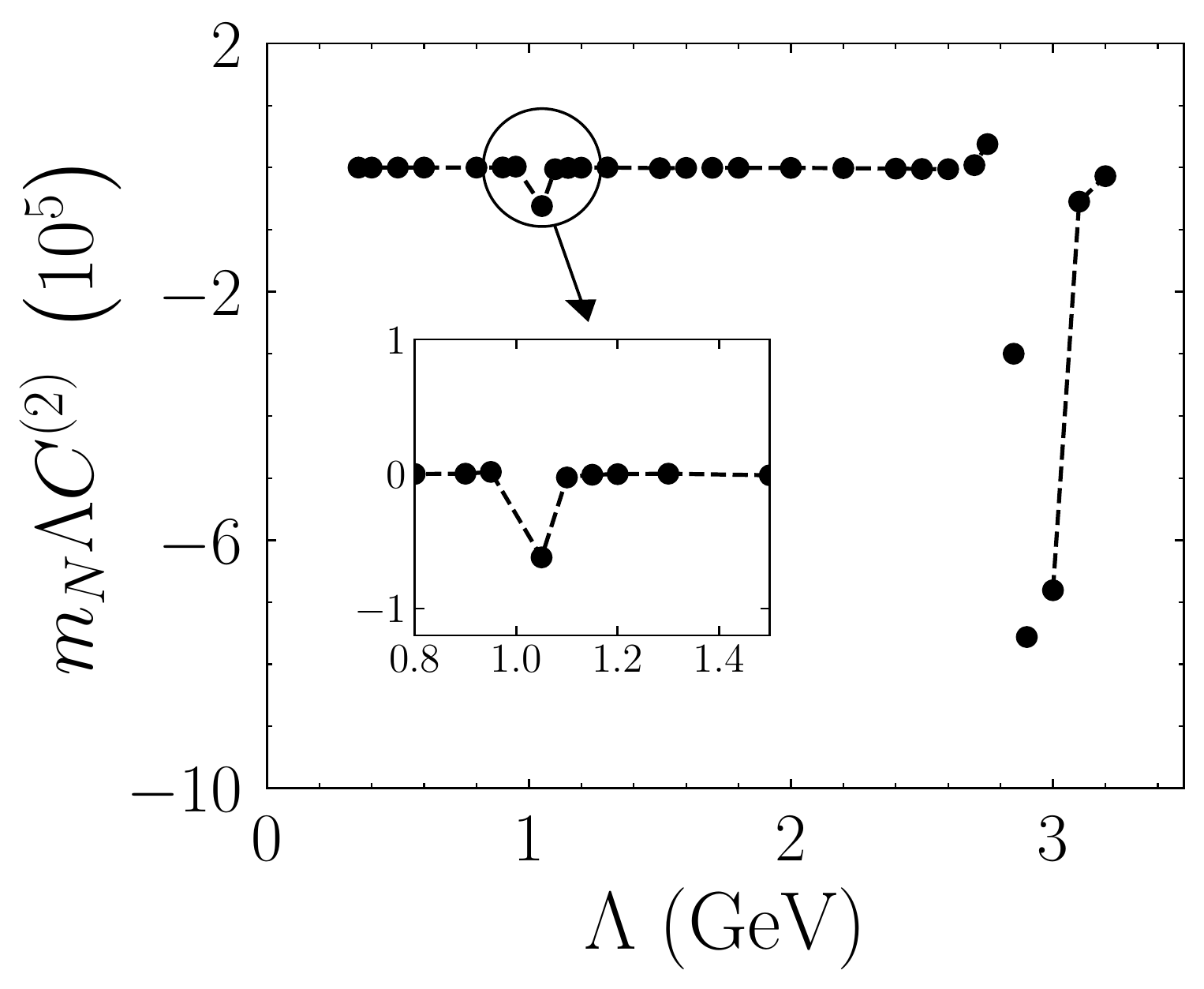}
    \includegraphics[scale=0.4]{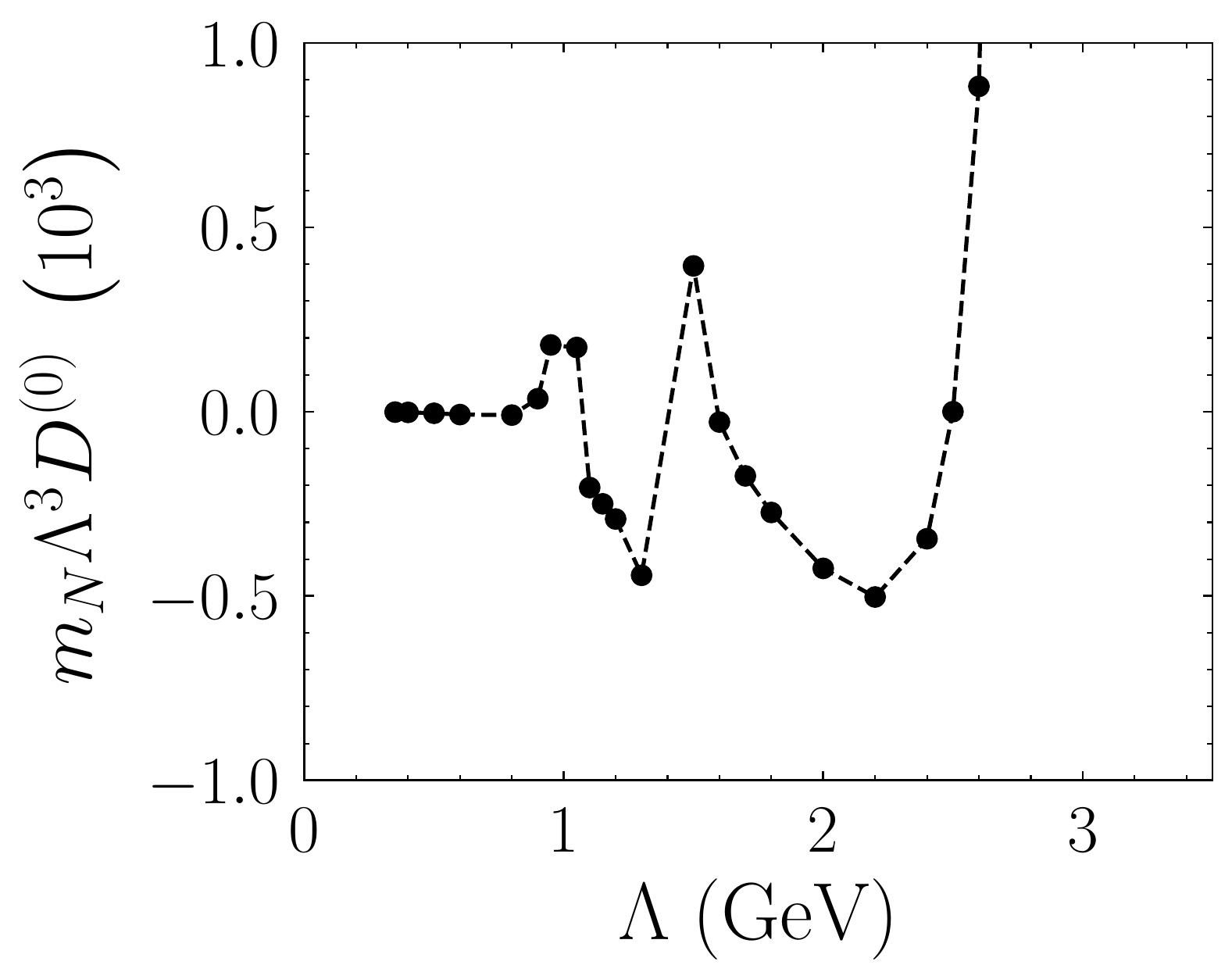}
    \includegraphics[scale=0.4]{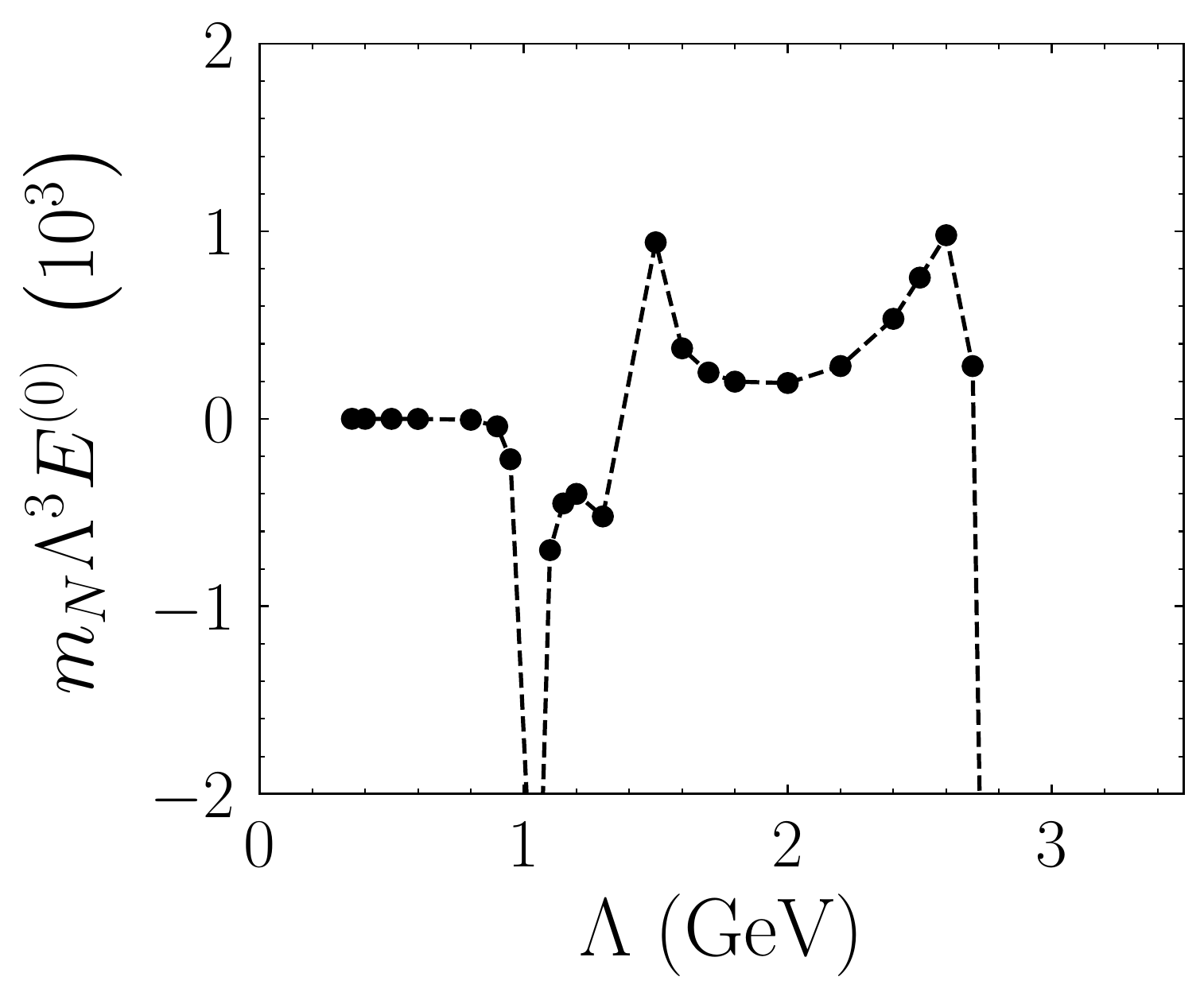}
    \caption{The LEC values of the {\NNLO} contact potentials as functions of $\Lambda$. 
    }
    \label{fig:LECsRunning}
\end{figure}

Much like the LO $\cs{3}{1}$ contact coupling showing limit-cycle running against $\Lambda$, the {\NNLO} contact LECs do too, as seen in Fig.~\ref{fig:LECsRunning} where dimensionless combinations of $\Lambda$ and $C^{(2)}$, $D^{(0)}$, and $E^{(0)}$ are shown: $m_N \Lambda C^{(2)}$, $m_N \Lambda^3 D^{(0)}$, and $m_N \Lambda^3 E^{(0)}$ with the nucleon mass $m_N = 939.0$ MeV. The {\NNLO} LECs appear to diverge near certain cutoff values. For instance, the value of $C^{(2)}$ increases by five orders of magnitude near $\Lambda = 1.0$ and $3.0$ GeV. This turns out to hinder numerical accuracy of the {\NNLO} correction---the incremental change from LO to {\NNLO}. In fact, the numerical accuracy of the {\NNLO} LEC values themselves near these cutoffs is questionable. An investigation on the mechanism driving these divergences is progress.

\begin{figure}
    \centering
    \includegraphics[scale=0.42]{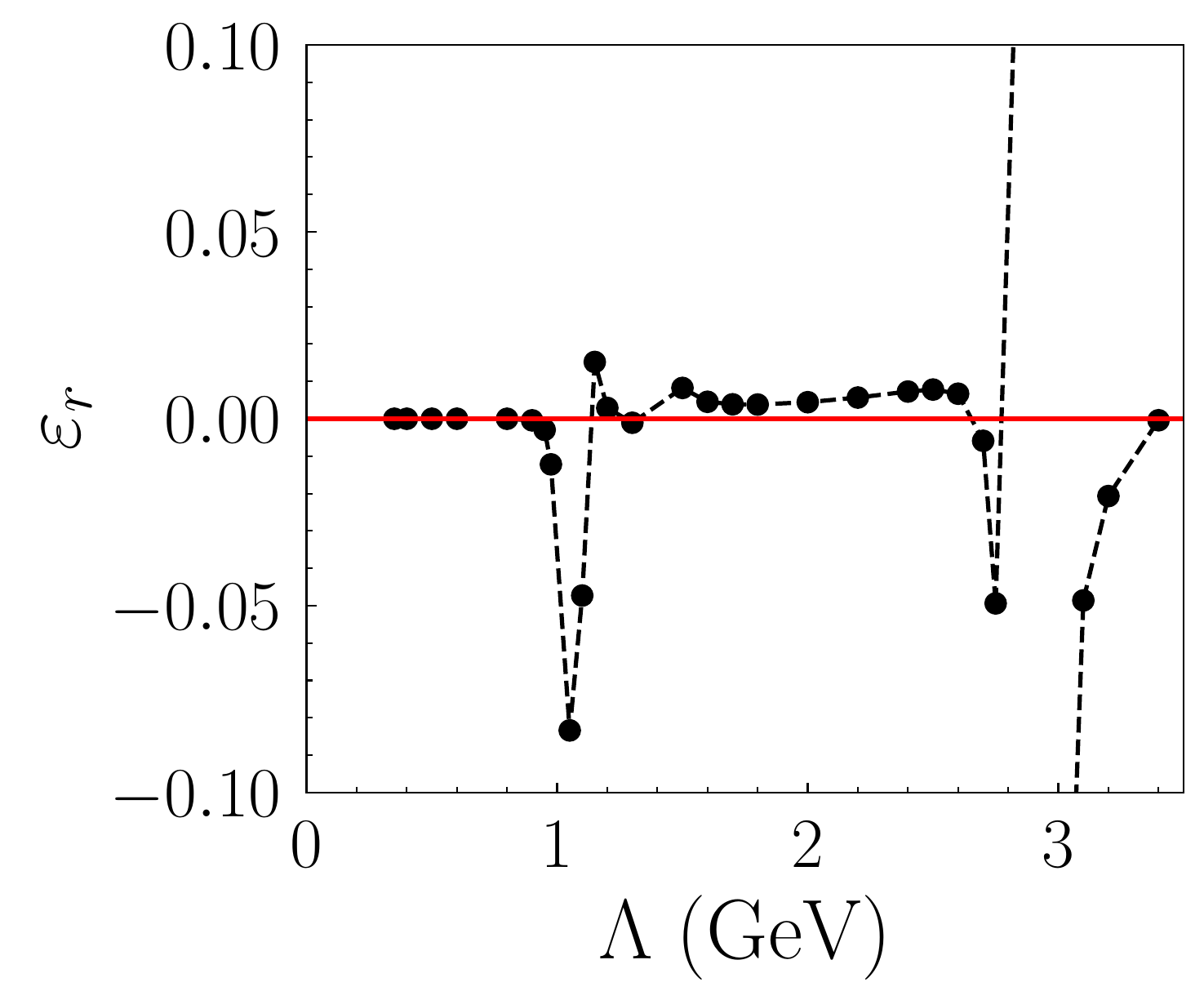}
    \caption{The ratio $\varepsilon_r \equiv (\varepsilon_\text{\NNLO} - \varepsilon_\text{emp})/(\varepsilon_\text{\NNLO} - \varepsilon_\text{LO})$ at $k = 150.12$ MeV as a function of $\Lambda$, where $\varepsilon_\text{LO}$ and $\varepsilon_\text{\NNLO}$ are LO and {\NNLO} values of the mixing angle $\varepsilon_1$, and $\varepsilon_\text{emp}$ is the empirical value from the Nijmegen partial wave analysis. 
    }
    \label{fig:e1_solve}
\end{figure}

We are thus forced to leave out certain cutoff windows in the calculations, with the criterion given as follows. At $k = 150.12$ MeV where the empirical mixing angle $\varepsilon_1$ is supposed to be reproduced, the relative error of the {\NNLO} correction to $\varepsilon_1$ must be within one percent. If our numerical calculations could be done with infinitely high accuracy for any cutoff values, the error would have been identically zero everywhere. As shown in Fig.~\ref{fig:e1_solve}, the error is larger than the one-percent criterion near $\Lambda = 1.0$ or $3.0$ GeV.

The deuteron wave function is obtained by solving the coupled homogeneous Lippmann-Schwinger equation in the $NN$ CM frame:
\begin{equation}
\begin{split}
  \psi_d(p) = \frac{1}{-B_d - \frac{p^{2}}{m_{N}}} \int dp' p^{\prime  2}\, V_{\cs{3}{1}\text{-}\cd{3}{1}}(p, p') \psi_d(p') \, ,
\end{split} \label{eqn:LSeqn}
\end{equation}
where $V_{\cs{3}{1}\text{-}\cd{3}{1}}(p, p')$ is a $2 \times 2$ matrix, $\psi_d(p)$ a column matrix with the reduced wave functions $u(p)/p$ and $w(p)/p$ as its components. $u(p)$ and $w(p)$ are normalized as
\begin{equation}
  \int_{0}^{\infty} dp \left[u^{2}(p) + w^{2}(p)\right] = 1  \, .
\end{equation}

It is useful to express the deuteron wave function in momentum space in terms of $u(p)$ and $w(p)$~\cite{Buck:1979ff}:
\begin{equation}
\psi_M(\vec{p}\,) \equiv \langle \vec{p}\, M | \psi_d \rangle = \sqrt{2}\, \pi \left[\frac{u(p)}{p} + \frac{w(p)}{p} \frac{1}{\sqrt{8}} \left(3\vec{\sigma}_1 \cdot \hat{p}\, \vec{\sigma}_2 \cdot \hat{p} - \vec{\sigma}_1 \cdot \vec{\sigma}_2\right)\right] \chi_{M}^{1}\, \eta_{0}^{0} \, , \label{eqn:DWF_uw}
\end{equation}
where $M = \pm1, 0$ is the $z$ component of the deuteron spin. Here the spin and isospin wave functions $\chi_{M}^1$ and $\eta_0^0$ can be expressed explicitly in the two-nucleon spin and isospin basis:
\begin{align}
\chi_1^1 ={}& \begin{pmatrix}
     1 \\
     0 
    \end{pmatrix}
    \begin{pmatrix}
     1 \\
     0 
    \end{pmatrix}  \, ,  \;
\chi_0^1 = \frac{1}{\sqrt{2}}\left[\begin{pmatrix}
     1 \\
     0 
    \end{pmatrix}
    \begin{pmatrix}
     0 \\
     1 
    \end{pmatrix}+\begin{pmatrix}
     0 \\
     1 
    \end{pmatrix}
    \begin{pmatrix}
     1 \\
     0 
    \end{pmatrix}\right]\, ,  \;
\chi_{-1}^1 = \begin{pmatrix}
     0 \\
     1 
    \end{pmatrix}
    \begin{pmatrix}
     0 \\
     1 
    \end{pmatrix}  \, ,  \\
\eta_0^0 ={}& \frac{1}{\sqrt{2}}\left[\begin{pmatrix}
     1 \\
     0 
    \end{pmatrix}
    \begin{pmatrix}
     0 \\
     1 
    \end{pmatrix}
    -\begin{pmatrix}
     0 \\
     1 
    \end{pmatrix}
    \begin{pmatrix}
     1 \\
     0 
    \end{pmatrix}\right]\, ,
\end{align}
and they are normalized so that
\begin{equation}
\left(\chi_{M'}^1\right)^\dagger \chi_{M}^1 = \delta_{M^\prime M}\, , \quad {\eta_0^0}^\dagger \eta_{0}^0  = 1\, . \nonumber
\end{equation}

We follow the method given in Ref.~\cite{Friar:1977xh, Schiavilla:2002fq} to kinematically boost the deuteron wave function to account for the fact that it is moving:
\begin{equation}
  \label{eq:boostedDWF}
  \psi_M(\vec{p}\,; \vec{v}\,) \simeq \left(1 - \frac{\vec{v}\,^2}{4}\right) \left[1 - \frac{1}{2} (\vec{v} \cdot \vec{p}\,) \left(\vec{v} \cdot \vec{\nabla}_{\!p}\right) - \frac{i}{4 m_N} \vec{v} \cdot (\vec{\sigma}_1 - \vec{\sigma}_2) \times \vec{p}\,\right] \psi_M(\vec{p}\,; 0) 
  \, ,
\end{equation}
where $\vec{v} \equiv \vec{p}_d/\sqrt{\vec{p}_d^{\;2} + m_d^2}$\,, with $m_{d}$ and $\vec{p}_d$ the deuteron mass and momentum. 

\section{Deuteron form factors\label{sec:formalism}}

The four-momentum transfer carried by the electron in the electron-deuteron elastic scattering is denoted by $q_\mu$ and the negative of its square conventionally by $Q^2$:
\begin{align}
   Q^2 \equiv -q_\mu q^\mu  \, .
     \label{eqn:defQ2}
\end{align}
The calculations will be preformed in the Breit frame, in which the initial and final momenta of the deuteron $\vec{p}_d$ and $\vec{p}\,'_d$ are equal in magnitude but opposite in their direction:
\begin{equation} \label{eqn:Breitframe}
     \vec{p}_d + \vec{p}_d^{\hspace{0.7mm}\prime} = 0 \, .
\end{equation}
Therefore, $q_0 = 0$, $Q^2 = \vec{q}\,^2$, and $\vec{p}_d = - \vec{q}/2$. It proves useful to define $\vec{K} \equiv\, \vec{p} + \frac{\vec{q}}{4}$.

In the Breit frame the charge, quadrupole, and magnetic form factors $G_C$, $G_Q$, and $G_M$ are related to the charge and current operators as follows:
\begin{align}
    G_{C}\left(Q^2\right) ={}& \frac{1}{3\, e} \sum_M \langle \psi_{M} | \rho | \psi_{M} \rangle
    \, , \label{eq:GC}   \\
  G_{Q}\left(Q^2\right) ={}& \frac{1}{2\, e\, \eta\, m_d^2} \left(\langle \psi_{M=0} | \rho | \psi_{M=0} \rangle - \langle \psi_{M=1} | \rho|\psi_{M=1} \rangle\right) \, ,    \label{eq:GQ}   \\
  G_{M}\left(Q^2\right) ={}& -\frac{1}{\sqrt{2\, \eta}\, e} \langle\psi_{M=1} | J^+ | \psi_{M=0} \rangle \, , \label{eq:GM}
\end{align}
where $J^{+} \equiv (J_1 + iJ_2)$, $\eta \equiv Q^2/(4m_d^2)$, and $e$ is the magnitude of the electron charge. 
The form factors are normalized so that
\begin{equation}
   G_{C}(0) = 1,\, G_{Q}(0) = Q_{d},\, G_{M}(0)=\mu_{d}\, \frac{m_{d}}{m_{N}}  \, ,   \nonumber
\end{equation}
where $Q_{d}$ is the deuteron electric quadrupole moment and $\mu_{d}$ the magnetic moment in units of nuclear magnetons. The charge radius can be expressed as the derivative of $G_C(Q^2)$ with respect to $Q^{2}$ at $Q^{2}=0$:
\begin{equation} \label{eq:r_d^2}
  \left\langle r_{d}^{2} \right\rangle \equiv -6 \left.\frac{d\, G_{C}}{d\, Q^{2}}\right|_{Q^2 = 0} \, .
\end{equation}
The experimental values of $Q_{d}$, $\mu_d$, and $r_d^2$ can be found in the literature: $Q_d = 0.2859(3)$ fm$^2$~\cite{Code:1971zz, Gambhir:1979zz}, $\mu_d = 0.8574$~\cite{Mohr:2015ccw}, and $r_d^2 = 4.537(51)$ fm$^2$~\cite{Sick:1998cvq}.

There are multiple low-energy scales in the problem: the magnitude of the momentum transfer $q \equiv |\vec{q}\,|$, the deuteron binding momentum $\gamma_d \simeq 47$ MeV, and the pion mass $m_\pi$. For the moment, we do not hierarchize further according to their relative size and simply catalog the first two orders of charge and current operators according to the power of these soft scales. 

Up to {\NNLO} in NDA, the long-range contributions to the deuteron charge, electric quadrupole, and magnetic form factors turn out to involve only one-body electromagnetic current operators~\cite{Ohtsubo:1970js, Chemtob:1971pu, Tsushima:1992nqz, Park:1995pn, Walzl:2001vb, Pastore:2008ui, Pastore:2009is, Kolling:2009iq, Pastore:2011ip, Kolling:2011mt, Krebs:2019aka}. Making use of this observation, we factor out the momentum-conserving delta function:
\begin{equation}
    \langle \vec{p}\,' | J_\mu(\vec{q}\,) | \vec{p}\, \rangle = (2\pi)^3 \delta^{(3)}\left(\vec{p}\,' - \vec{p} - \frac{\vec{q}}{2}\,\right) J_\mu\left(\vec{p}\,' , \vec{p} \, ; \vec{q}\, \right)\, .
    \label{eqn:deltaFunc}
\end{equation}
Unless noted otherwise, the charge and current operators are always expressed in the Breit frame. At LO they are given by~\cite{Pastore:2008ui, Pastore:2009is, Pastore:2011ip}
\begin{align}
\rho^{(0)} \left(\vec{p}\,' , \vec{p} \, ; \vec{q}\, \right) ={}& e \, ,  \label{eq:LOcharge_NFFCh}   \\
\vec{J}^{\,(0)}
  \left(\vec{p}\,' , \vec{p} \, ; \vec{q}\, \right) ={}& \frac{e}{2\, m_{N}}\left[2\vec{K} + i(1 + \kappa_s) \vec{\sigma} \times \vec{q} \,\right]\, ,     \label{eq:LOcurrent_NFFCh}  
\end{align}
with $\kappa_s = -0.12$ the isoscalar anomalous magnetic moment of the nucleon. The NLO charge operator vanishes, and the NLO current operator does not contribute because it is made of isovector terms proportional to $\bm{\tau}_1 \times \bm{\tau}_2$, whereas the deuteron is an isoscalar. 

At {\NNLO}, the deuteron charge operator can be split into two parts: one is the $1/m_N^2$ correction to the heavy-baryon limit charge operator at LO and the other is due to the structure of the nucleon~\cite{Pastore:2011ip}, thus scaling with $m_\pi^2$:
\begin{align}
    \rho_{\text{rel}}^{(2)}\left(\vec{p}\,' , \vec{p} \, ; \vec{q}\, \right) ={}& -\frac{e}{8\, m_N^2}\, \left(1 + 2\kappa_{s}\right) 
    \left(q^2 + 2i\, \vec{q} \cdot \vec{\sigma} \times \vec{K}\right)  \, ,    \label{eq:NNLOcharge_rel_NFFCh}  \\
  \rho_{\text{str}}^{(2)}\left(\vec{p}\,' , \vec{p} \, ; \vec{q}\, \right) ={}& -\frac{e}{6}\, \left\langle r_{ES}^2 \right\rangle\, q^{2}  \, ,  \label{eq:NNLOcharge_structure_NFFCh}
\end{align}
where $\left\langle r_{ES}^2 \right\rangle = (0.777\, \text{ fm})^2$ is the isoscalar charge radius of the nucleon. A similar categorization applies to the {\NNLO} current operator~\cite{Pastore:2008ui, Pastore:2009is}:
\begin{align}
   \vec{J}_{\text{rel}}^{\,(2)}\left(\vec{p}\,' , \vec{p} \, ; \vec{q}\, \right) ={}& - \frac{e}{8\, m_{N}^{3}} \Big{[}2 \left(\vec{K}^{2} + \frac{q^2}{4} \right) \left(2\vec{K} + i\vec{\sigma} \times \vec{q}\, \right)   
   + \vec{K} \cdot \vec{q}\, \left(\vec{q} + 2i \vec{\sigma} \times \vec{K}\, \right) \Big{]}      \nonumber \\
 &- \frac{i\, e\, \kappa_s} {8\, m_{N}^{3}} \Big{[}\vec{K} \cdot \vec{q}\, \left(4\vec{\sigma} \times \vec{K} - i\vec{q}\, \right)    
  - \left(2i \vec{K} - \vec{\sigma} \times \vec{q}\, \right) \frac{q^2}{2}  \nonumber \\
  &\qquad\qquad + 2\vec{K} \times \vec{q}\,\, \vec{\sigma} \cdot \vec{K} \Big{]}   \, ,      
  \label{eq:NNLOcurrent_rel_NFFCh}   \\
  \vec{J}_{\text{str}}^{\,(2)}\left(\vec{p}\,' , \vec{p} \, ; \vec{q}\, \right) ={}& -\frac{e}{12\, m_{N}} \left[2 \left\langle r_{ES}^2 \right\rangle q^2 \vec{K} + i(1 + \kappa_s)  \left\langle r_{MS}^2 \right\rangle q^2
    \vec{\sigma} \times \vec{q}\, \right]   \, , 
    \label{eq:NNLOcurrent_structure_NFFCh}    
\end{align}
where $\left\langle r_{MS}^2 \right\rangle = (0.707\, \text{ fm})^2$ is the isoscalar magnetic radius of the nucleon.

As will be discussed in Sec.~\ref{sec:results} and also suggested in Ref.~\cite{PavonValderrama:2014zeq}, there is evidence that the contact current operators responsible for $G_M$ are enhanced. We will need the lowest-dimension, non-vanishing one~\cite{Pastore:2009is, Kolling:2011mt, Kolling:2012cs, Phillips:2016mov}:
\begin{equation}
\langle \vec{p}\,' | \vec{J}_{\text{ct}} | \vec{p}\, \rangle
  = i\, e\, L_2 \left[\left(\vec{\sigma}_1 + \vec{\sigma}_2\right) \times \vec{q}\,\right]   \, ,
  \label{eqn:NNNLOcurrent_ct_NFFCh}
\end{equation}
where $L_2$ is the LEC of the operator. The corresponding Lagrangian operator is given by~\cite{Kaplan:1998sz, Kolling:2011mt, Kolling:2012cs}
\begin{align}
    \mathcal{L}_{\, \gamma N} = -e\, L_2 \left(\vec{\nabla} \times \vec{A}\, \right) \cdot \left(N^\dag\, \vec{\sigma}\, N N^\dag N\right)  \, ,
\end{align}
where $\vec{A}$ is the external electromagnetic vector potential.

For one-body operators, the matrix elements involved in the deuteron form factors have a generic integral form:
\begin{equation}
\begin{split}
     \langle \psi_{M^\prime} | J_\mu | \psi_{M} \rangle = \int \frac{d^{3}p}{(2\pi)^{3}} \psi_{M^\prime}^{\dagger} \left(\vec{p}\,' ; \vec{v}\,\right) J_\mu\left(\vec{p}\,' , \vec{p} \, ; \vec{q}\, \right)\, \psi_{M} \left(\vec{p}\,; -\vec{v}\,\right) \, , \label{eq:onenucleon_matele}
\end{split}
\end{equation}
where $\vec{p}\,' = \vec{p} + \vec{q}/2$, as required by the momentum-conserving delta function in Eq.~\eqref{eqn:deltaFunc}, and $\vec{v} = \vec{q}/\sqrt{q^{\,2} + 4m_d^2}$. The corrections to $G_C$ and $G_Q$ by $\rho^{(2)}_\text{str}$ factorize into products involving their LO values:
\begin{align}
    G_{C}\left(Q^2| \rho_\text{str}^{(2)}\right) ={}& -\frac{1}{6} \langle r_{ES}^2 \rangle Q^2\, G_{C}\left(Q^2| \rho^{(0)}\right) \, ,  \\
    G_{Q}\left(Q^2| \rho_\text{str}^{(2)}\right) ={}& -\frac{1}{6} \langle r_{ES}^2 \rangle Q^2\, G_{Q}\left(Q^2| \rho^{(0)}\right) \, .      
\end{align}
Using Eqs.~\eqref{eqn:DWF_uw}, \eqref{eq:boostedDWF}, and the expressions for other charge and current operators  [Eqs.~\eqref{eq:LOcharge_NFFCh}--\eqref{eq:NNLOcurrent_structure_NFFCh}], we can reduce the integral to a two-dimensional version: one is over $|\vec{p}\,|$ and the angle between $\vec{p}$ and $\vec{q}$. For instance, $G_C$ generated by the LO charge operator $\rho^{(0)}$ is given by
\begin{equation}
    G_{C}\left(Q^2 | \rho^{(0)} \right) = \frac{1}{2} \int_{0}^{\infty} dp \int_{-1}^{1} dz\, 
    \frac{p}{p'} \left[u(p') u(p) + P_2\left(x\right) w(p') w(p)\right] \, ,     
    \label{eqn:LOGC_integral}
\end{equation}
where $z = \hat{p} \cdot \hat{q}$,
\begin{align}
    p' \equiv{}& |\vec{p}\,'| = \left(p^{2} + pqz + \frac{q^{2}}{4} \right)^\frac{1}{2} \, ,\\
    x \equiv{}& \hat{p}'\cdot \hat{p} = \frac{1}{p'}\left(p + \frac{1}{2} qz\right) \, ,     
\end{align}
and $P_2(x)$ is the second-degree Legendre polynomial. Similarly, $G_Q$ induced by $\rho^{(0)}$ can be expressed as
\begin{equation}
\begin{split}
     G_{Q}\left(Q^2| \rho^{(0)} \right) ={}& -\frac{3}{4\,q^2} \int_{0}^{\infty} dp \int_{-1}^{1} dz\, \frac{p}{p'} \Big{\{} 2\sqrt{2}P_2(z) u(p')w(p) \\
     &+ 2\sqrt{2} P_2(y) w(p')u(p) 
     - \left[P_2(x) + P_2(y) + P_2(z) - 1\right] w(p')w(p) \Big{\}}  \, , 
\label{eqn:LOGQ_integral}
\end{split}
\end{equation}
where 
\begin{equation}
    y \equiv \hat{p}\,' \cdot \hat{q} = \frac{1}{p'} \left(pz + \frac{1}{2} q\right) \, .
\end{equation}

Due to the angular-momentum coupling coefficients, the expressions become increasingly complex and lengthy for current operators and higher-order charge operators. We relegate the explicit formulas for integrals involving $\vec{J}^{\,(0)}$, $\rho^{(2)}_{\text{rel}}$, $\vec{J}^{\,(2)}_{\text{rel}}$, $\vec{J}^{\,(2)}_{\text{str}}$, and $\vec{J}_\text{ct}$ to the Appendix. The contact current operator, $\vec{J}_\text{ct}$ defined in Eq.~\eqref{eqn:NNNLOcurrent_ct_NFFCh}, generates a matrix element involving integrals no more complex than these one-body operators, despite its two-body nature.

The fact that the deuteron is not at rest in the Breit frame needs to be accounted for at {\NNLO}. According to Eq.~(\ref{eq:boostedDWF}) the boost-corrected matrix elements of the LO charge and current operators can be written as~\cite{Phillips:2003jz, Filin:2020tcs}
\begin{equation}
  \left\langle \psi_{M'} \left| J_{\mu}^{(0)} \right| \psi_{M} \right\rangle = \int \frac{d^{3}p}{(2\pi)^{3}} \psi_{M'}^{\dag}\left(\vec{p} + \frac{\vec{q}}{2\sqrt{1 + \eta}};0\right) J_{\mu}^{(0)}\left(\vec{p}\,' , \vec{p} \, ; \vec{q}\, \right) \psi_{M}
    \left(\vec{p}\,;0\right) \, .
    \label{eqn:boost_mar_ele}
\end{equation}
In keeping with the principle of calculating higher-order corrections perturbatively, we will use the LO wave function in Eq.~\eqref{eqn:boost_mar_ele}, expanding the resulting matrix element in $\eta$ and retaining the first term in the series. We note that this boost correction is mainly kinematic, and interaction-dependent relativistic corrections do not contribute. For more explanation, we refer to Refs.~\cite{Friar:1977xh, Schiavilla:2002fq, Lorce:2022jyi}. 

In dealing with {\NNLO} contributions from the {\NNLO} charge and current operators $\rho^{(2)}$ and $\vec{J}^{\, (2)}$, we must use the LO deuteron wave function. We are left with the question of how to correct the deuteron wave function by the {\NNLO} $\cs{3}{1}\text{-}\cd{3}{1}$ chiral potential. To treat the {\NNLO} potential as perturbation, we define a potential with an auxiliary variable $x$:
\begin{equation}
    V(x)=V^{(0)} + x V^{(2)}\, .
\end{equation}
We then calculate, say, the deuteron charge form factor via Eqs.~\eqref{eqn:LSeqn} and \eqref{eqn:LOGC_integral}, obtaining a generating function $G_C(Q^2|\rho^{(0)}; x)$. The desired correction due to $V^{(2)}$, denoted by $G_C^{(2)}\left(Q^2|\rho^{(0)}\right)$, is the coefficient of the $x$ term in the Taylor expansion of $G_C(Q^2|\rho^{(0)}; x)$:
\begin{align}
    G_C(Q^2|\rho^{(0)}; x) = G_C\left(Q^2|\rho^{(0)}\right) + x\, G_C^{(2)}\left(Q^2|\rho^{(0)}\right) + \mathcal{O}(x^2) \, .
\end{align}

To summarize, we note that the subleading corrections will be calculated in strict perturbation theory. The NLO corrections to the form factors considered here all vanish. The {\NNLO} corrections fall into four categories: ``Relativistic''--- the matrix elements of $\rho^{(2)}_\text{rel}$ and $\vec{J}^{\, (2)}_\text{rel}$ between the LO wave functions; ``Structure''--- driven by $m_\pi^2$, the matrix elements of $\rho^{(2)}_\text{str}$ and $\vec{J}^{\, (2)}_\text{str}$ between the LO wave functions; ``Boost''--- the correction contributed by the boosted LO wave function; and ``Potential''--- the correction to the deuteron wave function due to the {\NNLO} chiral potential.

\section{Results and Discussions\label{sec:results}}

\begin{figure}
\centering
\includegraphics[scale=0.35]{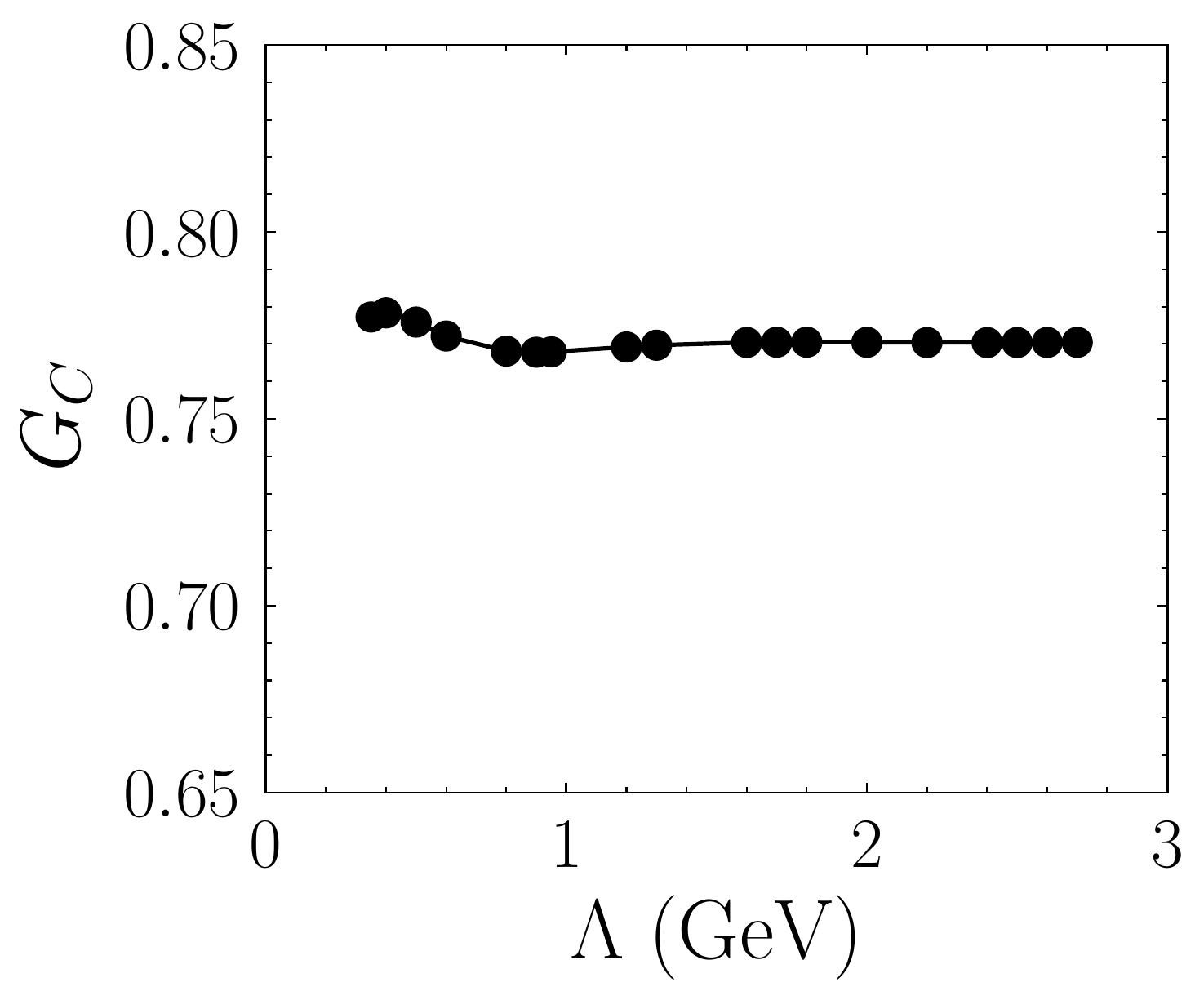}
\includegraphics[scale=0.35]{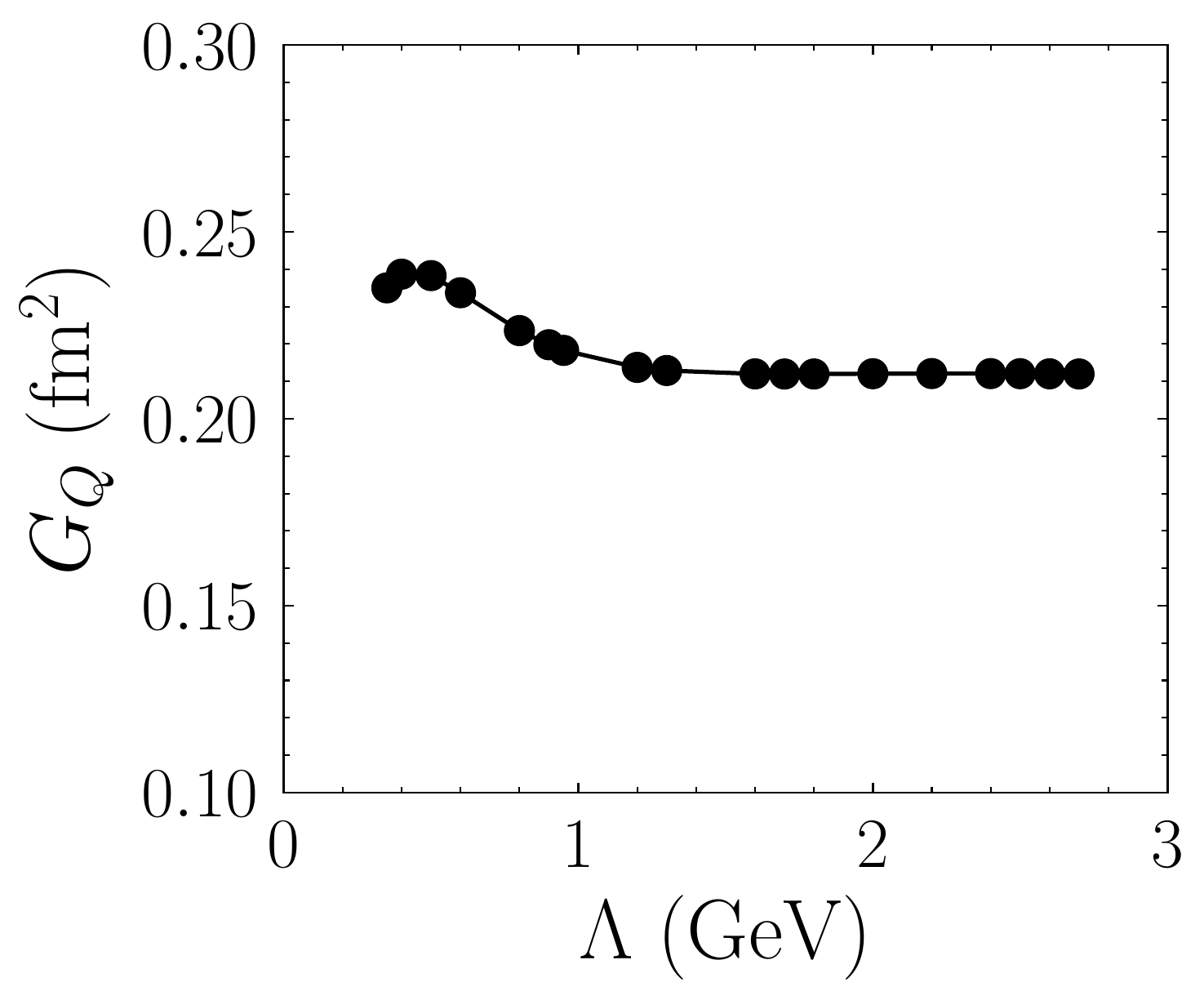}
\includegraphics[scale=0.35]{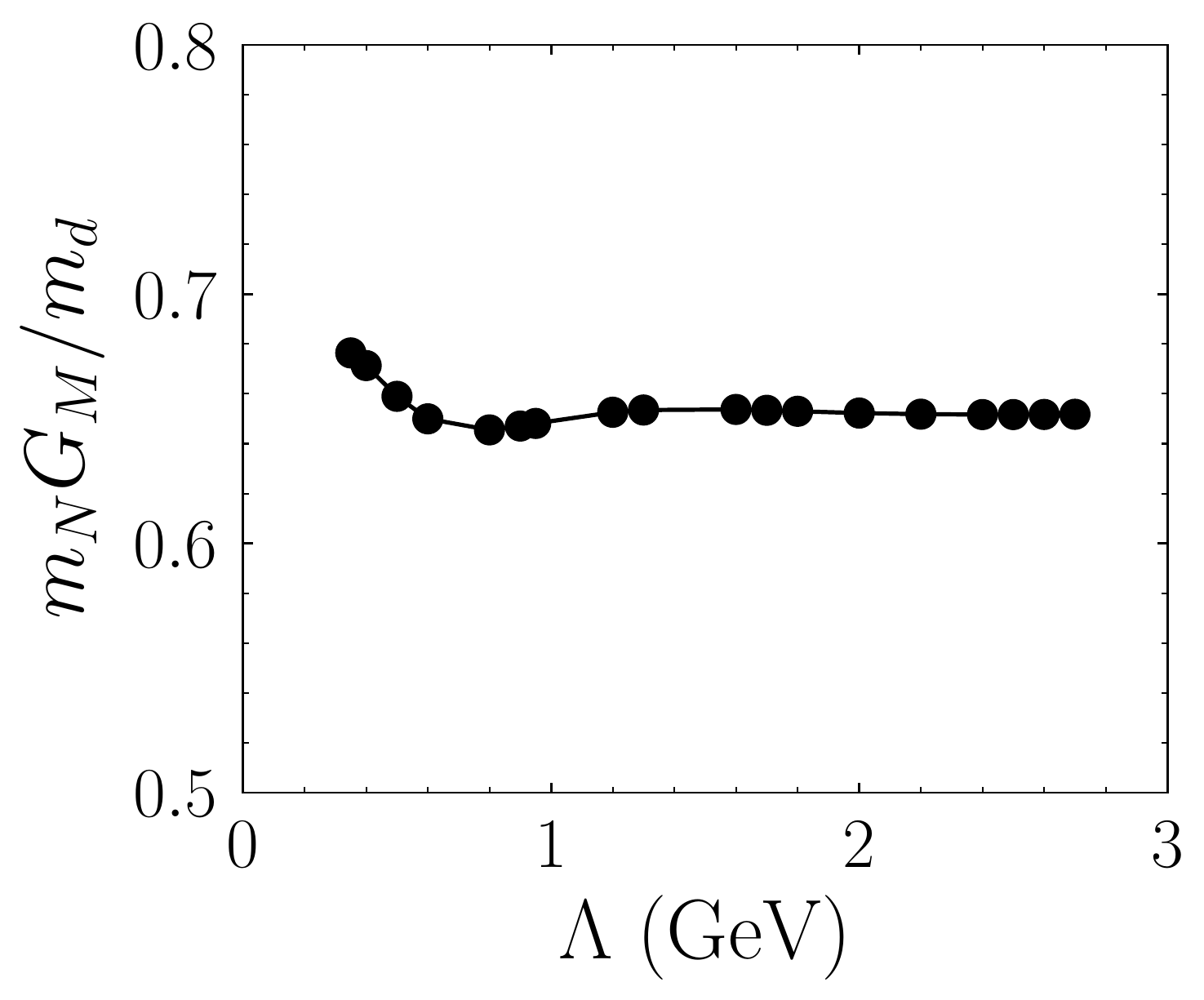}
\caption{The LO deuteron charge, quadrupole, and magnetic form factors at $q = m_\pi$ as functions of the UV cutoff value $\Lambda$.} 
\label{fig:GCGQGM_LOcutoff}
\end{figure}

We start by examining the cutoff variation of the deuteron form factors at LO and {\NNLO}. The LO form factors as functions of the cutoff value $\Lambda$ are shown in Fig.~\ref{fig:GCGQGM_LOcutoff} where the momentum transfer $q = m_\pi$ is chosen as a representative kinematic point. The cutoff-independent results at LO confirm the finding of Ref.~\cite{Valderrama:2007ja}.

\begin{figure}
\centering
\includegraphics[scale=0.35]{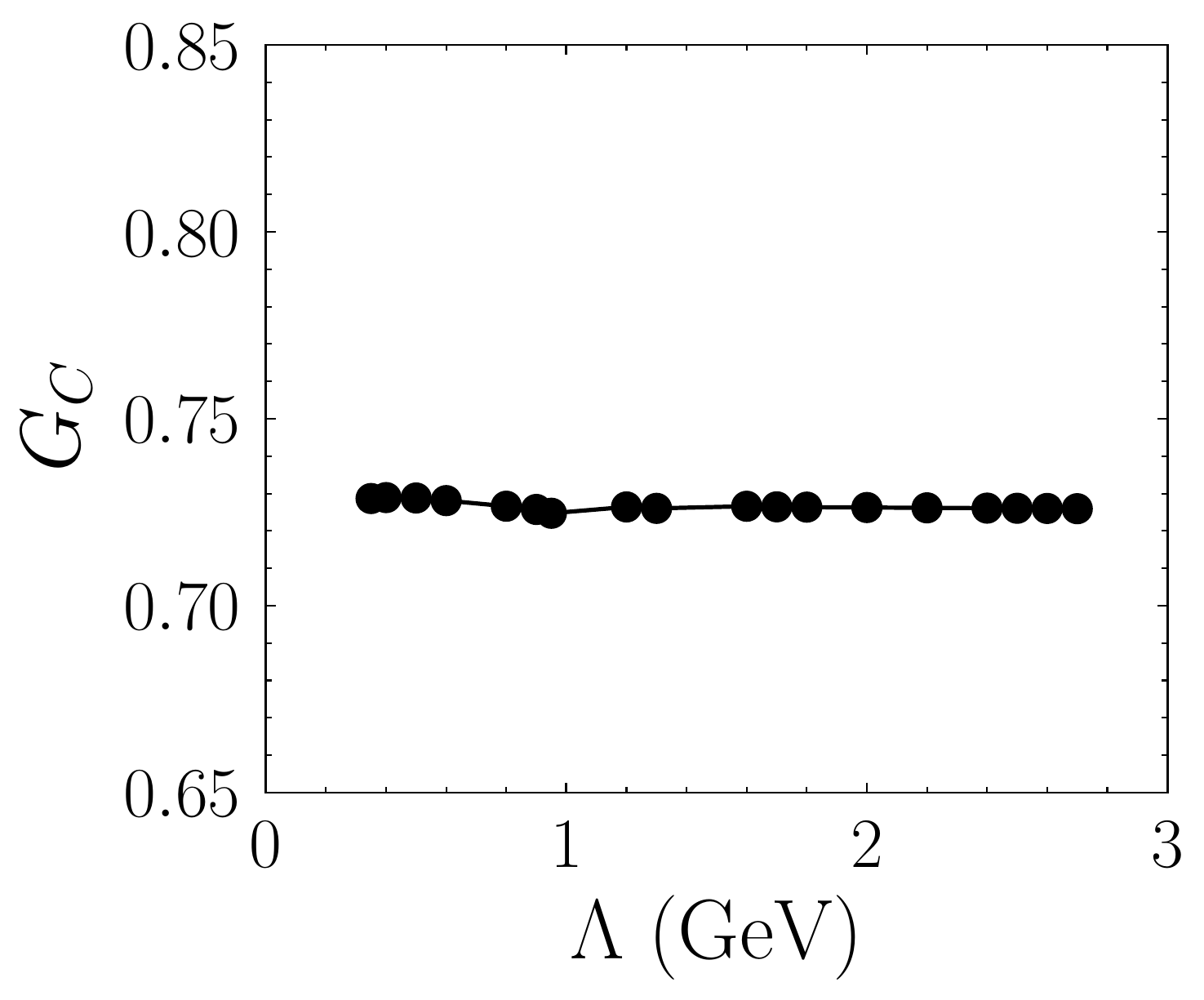}
\includegraphics[scale=0.35]{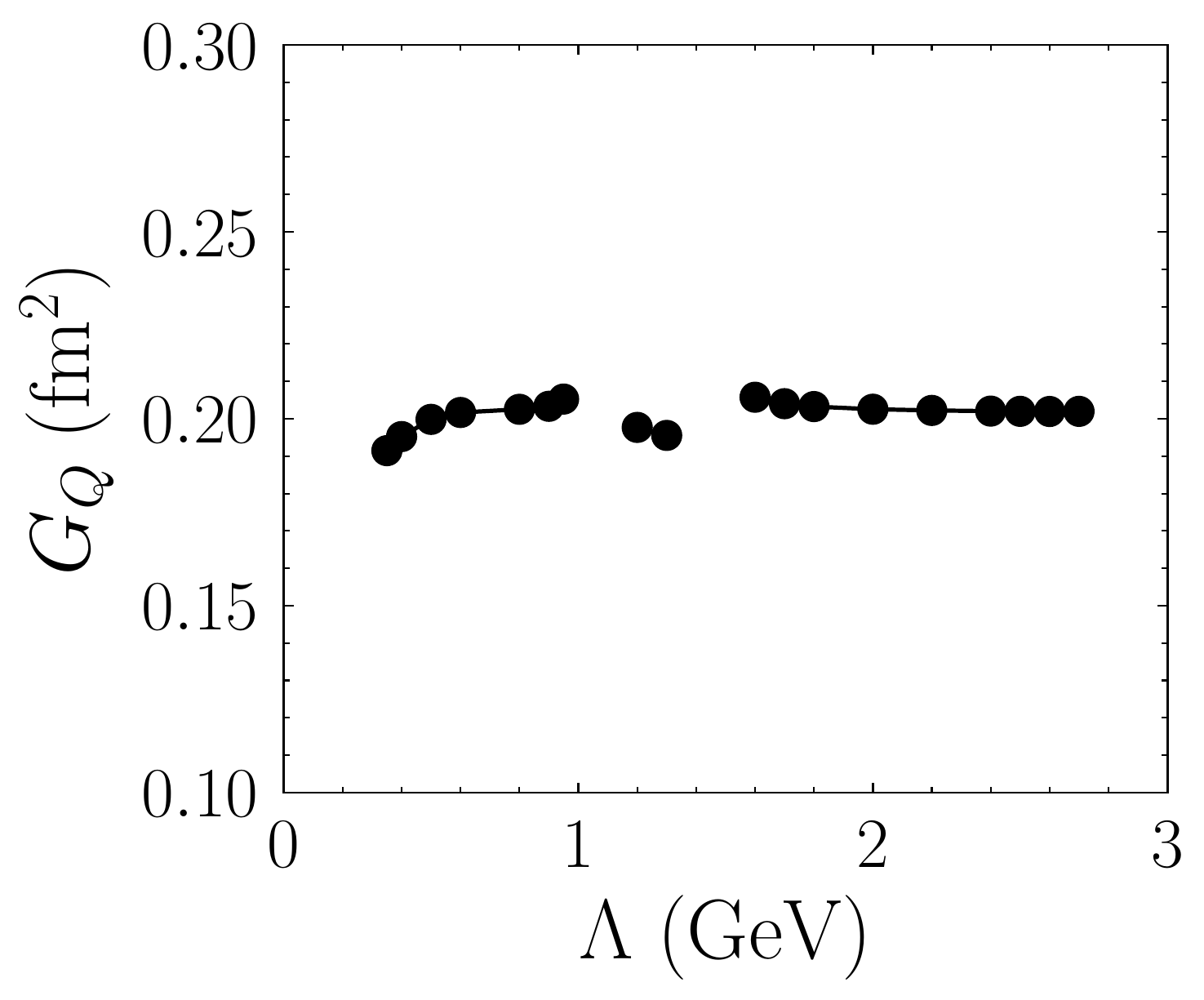}
\includegraphics[scale=0.35]{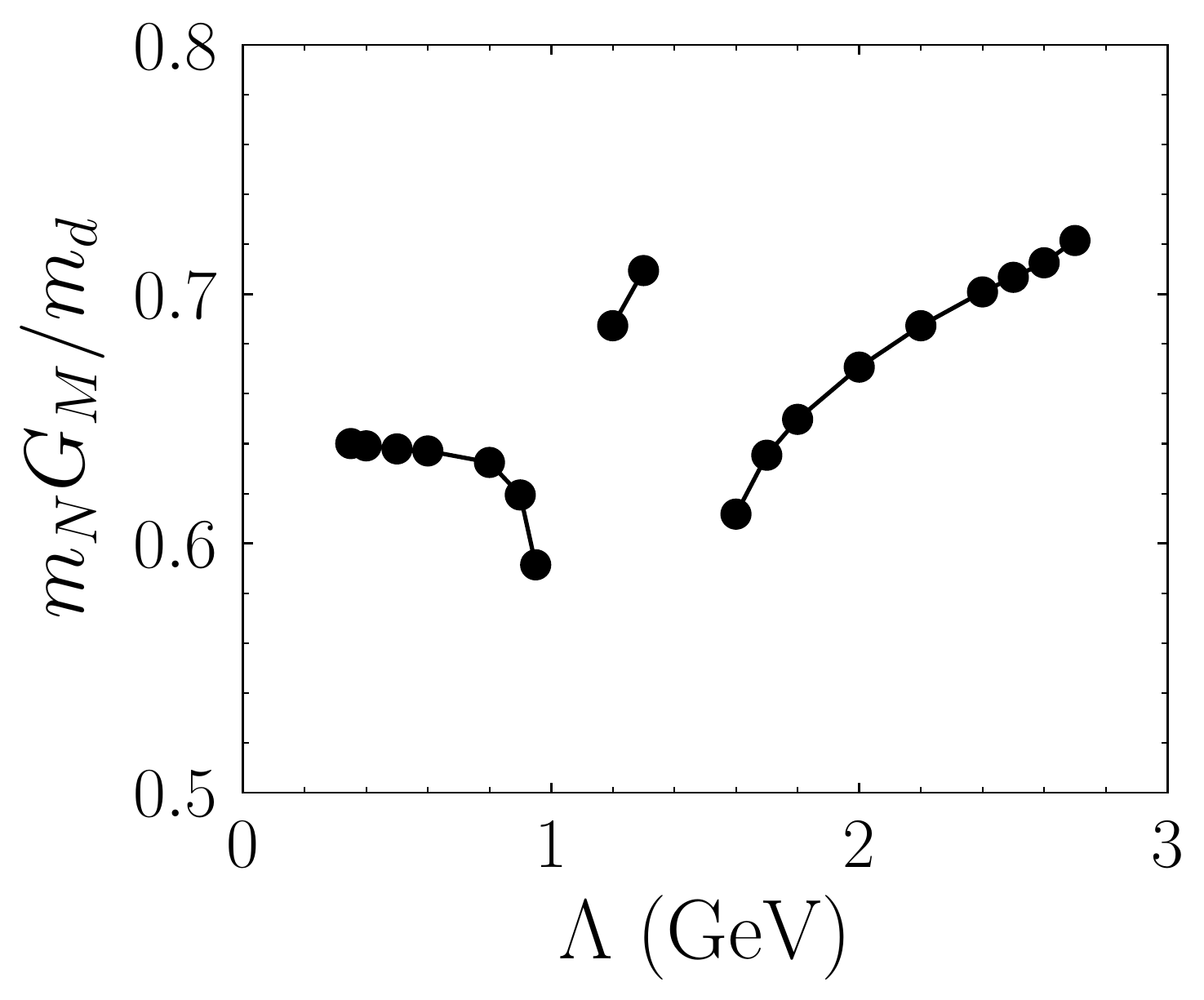}
\caption{The {\NNLO} deuteron charge, quadrupole, and magnetic form factors as functions of the UV cutoff value $\Lambda$ for $q = m_\pi$.} 
\label{fig:GCGQGM_NNLOcutoff}
\end{figure}

The cutoff-variation plots for {\NNLO} are shown in Fig.~\ref{fig:GCGQGM_NNLOcutoff}. $G_C$ and $G_Q$ converge well for large values of $\Lambda$. As a result, there is no motivation from the RG perspective to modify the NDA counting of the charge operators at this order. However, $G_M$ does not converge with respect to $\Lambda$. There appears to be some sort of limit-cycle-like oscillation between $\Lambda \simeq 900$ and 1600 MeV, but the numerical difficulty in this cutoff window unfortunately prevents us from studying more closely (see Sec.~\ref{sec:pot-dwf}). From $\Lambda = 1600$ MeV onward the {\NNLO} $G_M$ shows about $15\%$ variation in comparison with its LO value (see Fig.~\ref{fig:GCGQGM_LOcutoff}). Assuming that the most relevant soft scale is $m_\pi$ rather than $\gamma_d$ and that the breakdown scale is the delta-nucleon mass splitting $\delta \simeq 300$ MeV, we make a highly conservative estimation of theoretical uncertainty expected of the {\NNLO}: $(m_\pi/\delta)^3 \simeq 10\%$. Therefore, even if we ignore the fact that the {\NNLO} $G_M$ tends to increase linearly at high cutoff values in the plot, $15 \%$ is still quite a large variation to be acceptable uncertainty at {\NNLO}. 

\begin{figure}
\centering
\includegraphics[scale=0.35]{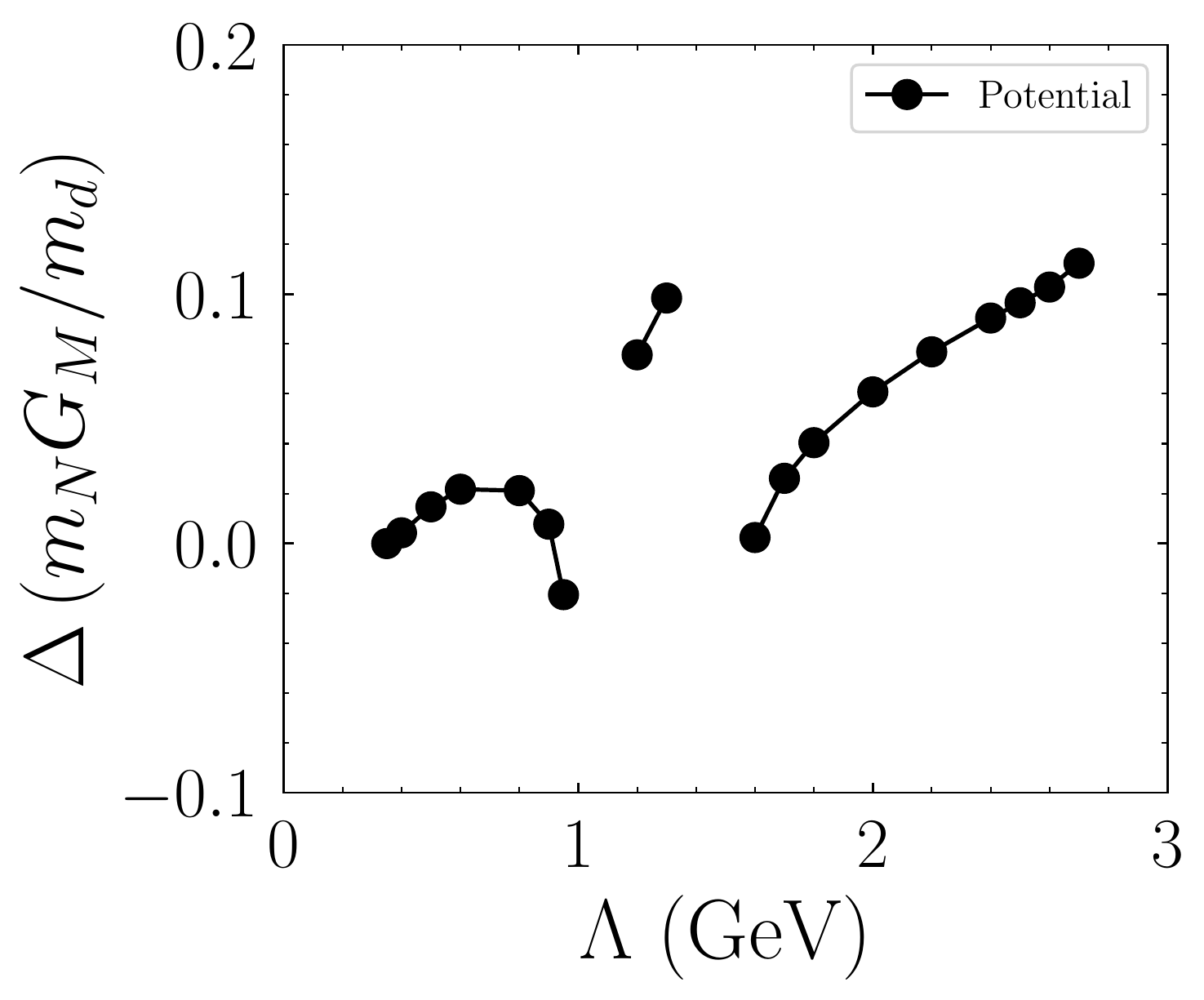}
\includegraphics[scale=0.35]{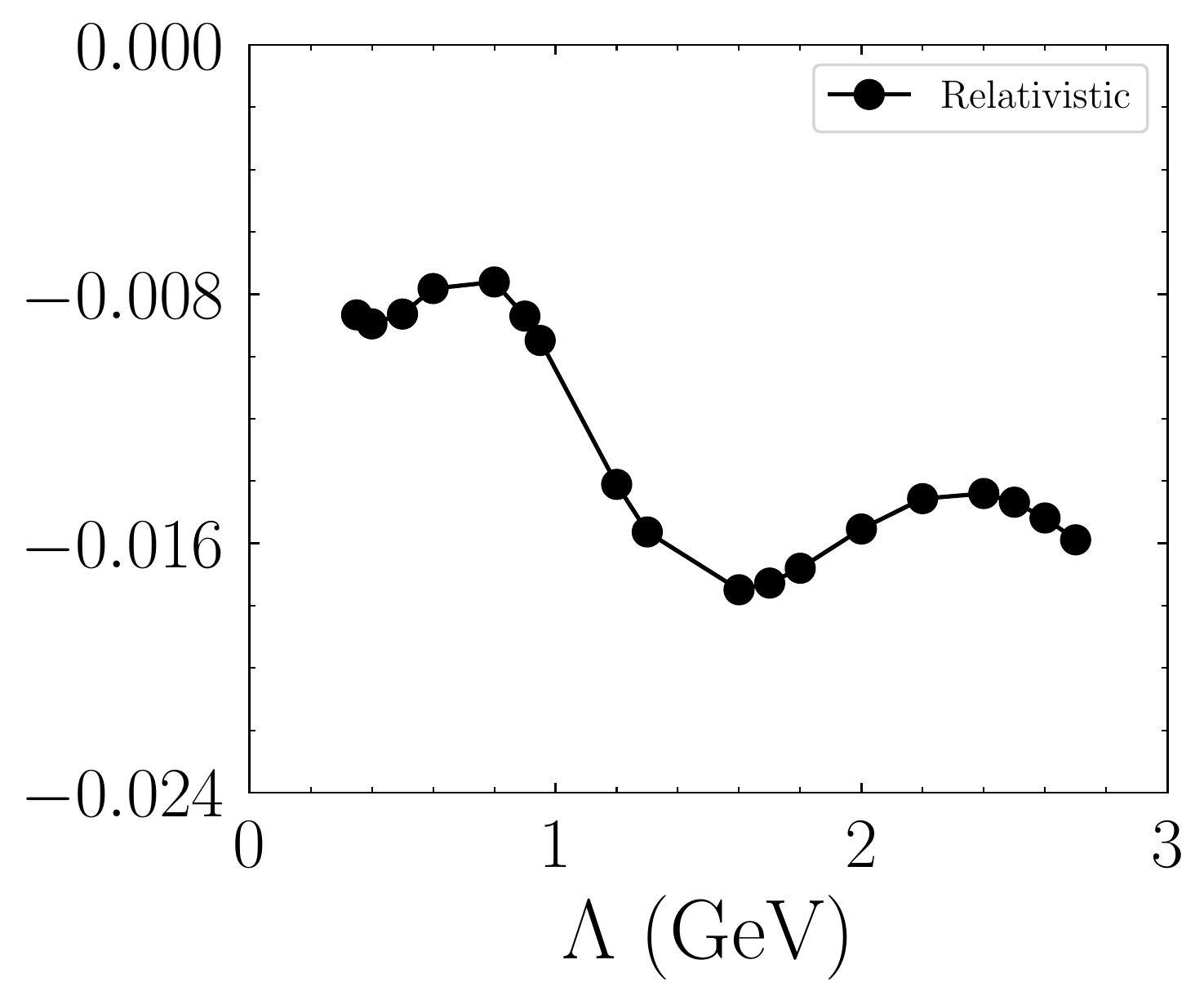}
\includegraphics[scale=0.35]{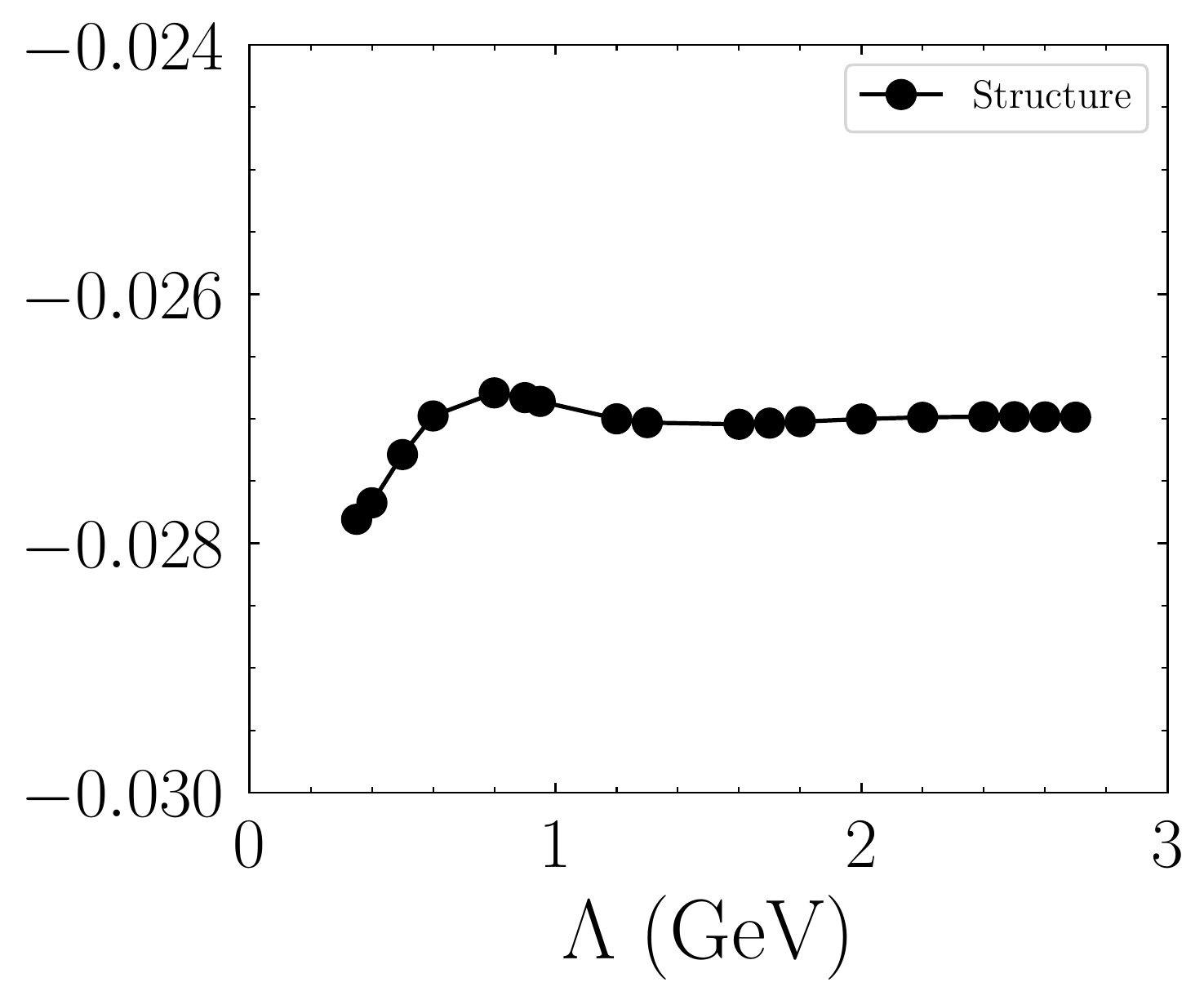}
\caption{The {\NNLO} correction to $G_M$ as a function of $\Lambda$ at $q = m_\pi$ before renormalization. The legends specify the source driving the correction. See the text for more explanation.} 
\label{fig:GM_cutoff_breakdown}
\end{figure}

To understand better this cutoff variation, we first notice that the potential contribution dominates its composition, as indicated in Fig.~\ref{fig:GM_cutoff_breakdown} where the {\NNLO} $G_M$ is broken down to the four classes that are cataloged at the end of Sec.~\ref{sec:formalism}. The size of the boost correction is only about $0.1 \%$ of the LO value, and it converges very rapidly, hence it is not shown in Fig.~\ref{fig:GM_cutoff_breakdown}. Because the structure correction stabilizes with respect to $\Lambda$, we know consequently that it is $\gamma/\Lambda$ and/or $q/\Lambda$ rather than $m_\pi/\Lambda$ that drives the cutoff variation of $G_M$ at {\NNLO}.

\begin{figure}
\centering
\includegraphics[scale=0.4]{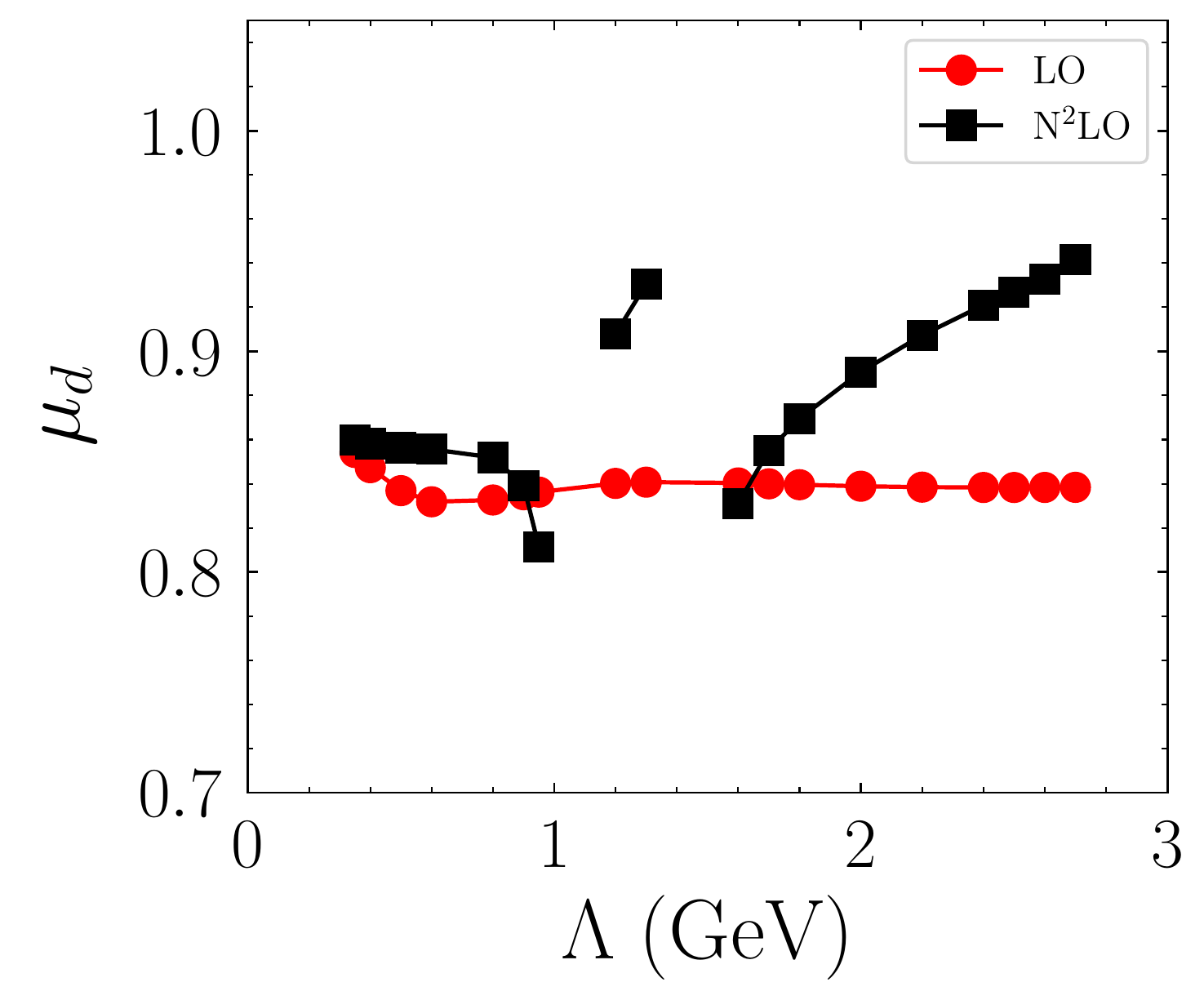}
\caption{The cutoff variation of $\mu_d$ at LO and {\NNLO} before renormalization.}
\label{fig:mu_d_total_cutoff}
\end{figure}

Second, we minimize the influence of $q$ and look at the deuteron magnetic moment $\mu_d$ which is defined at the limit $q^2 \to 0$. $\mu_d$ as a function of $\Lambda$ is plotted in Fig.~\ref{fig:mu_d_total_cutoff}. The uncertainty expected of the {\NNLO} potential correction to $\mu_d$ is controlled by $(\gamma_d/\delta)^3 \simeq 0.4 \%$, but this is much smaller than the ten-percent level variation demonstrated in Fig.~\ref{fig:mu_d_total_cutoff}. Therefore, we conclude that the contact two-body current operator \eqref{eqn:NNNLOcurrent_ct_NFFCh} must appear at {\NNLO} in order to cancel the rather large cutoff variation. 

\begin{figure}
\centering
\includegraphics[scale=0.4]{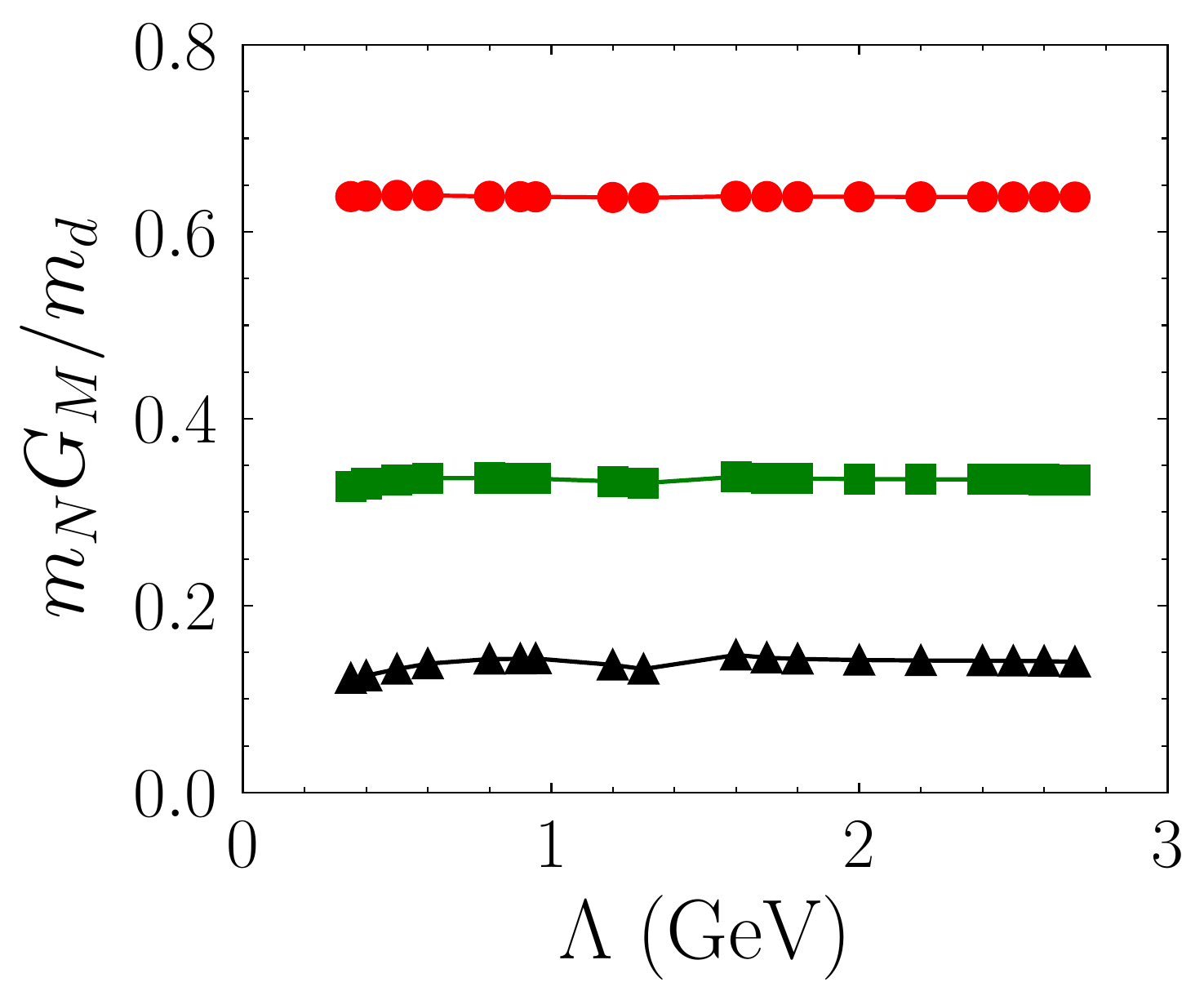}
\caption{
The cutoff variation of $G_M$ at {\NNLO} after renormalization. The solid circles, squares, and triangles correspond to $q = m_\pi$, $2m_\pi$, and $3m_\pi$.}
\label{fig:GMCTVsCutoff}
\end{figure}

We then study whether $\vec{J}_\text{ct}$ indeed renormalizes $G_M$. To determine $L_2$, we use the experimental value of $\mu_d$. The renormalized {\NNLO} $G_M$, which is the sum of contributions from {\NNLO} one-body operators and $\vec{J}_\text{ct}$, at various $q$ are then predicted for $\Lambda$, as plotted in Fig.~\ref{fig:GMCTVsCutoff}. After renormalization, cutoff independence is clearly achieved. 

According to NDA, the first non-vanishing, contact charge and current operators  appear at N$^5$LO and {\NNNLO}~\cite{Phillips:2003jz, Phillips:2006im}, respectively, in our notation of power counting.  Reference~\cite{PavonValderrama:2014zeq} proposed, however, that these contact operators appear at lower order, N$^{4.5}$LO and N$^{2.5}$LO. The RG analysis therein was based on the short-range behavior of the deuteron wave function generated by the LO chiral potential, Eqs.~\eqref{eqn:OPEpot} and \eqref{eqn:LOVS}. Although their approach is not identical to ours, we agree with their conclusion in the sense that $\vec{J}_\text{ct}$ is enhanced somewhat compared with its NDA estimation. 

In perturbative-pion ChEFT where the breakdown scale is smaller than nonperturbative-pion ChEFT, $L_2$ is expected to take different values. It is nevertheless found to be enhanced over NDA~\cite{Kaplan:1998sz}. Based on the assumption of nontrivial infrared fixed point of LECs and backed by analytic calculations of renormalization-scale dependence in power divergence subtraction scheme, the value of $L_2$ is found to scale as $Q^{-2}$. This puts the two-body contact current $\vec{J}_\text{ct}$ at NLO, two orders lower than NDA. Besides the magnetic one, the deuteron form factors of other probes, interesting to studies of fundamental symmetries, have been shown to be enhanced to various extents~\cite{Savage:1999cm, deVries:2011re, Mereghetti:2013bta}.

\begin{figure}
\centering
\includegraphics[scale=0.4]{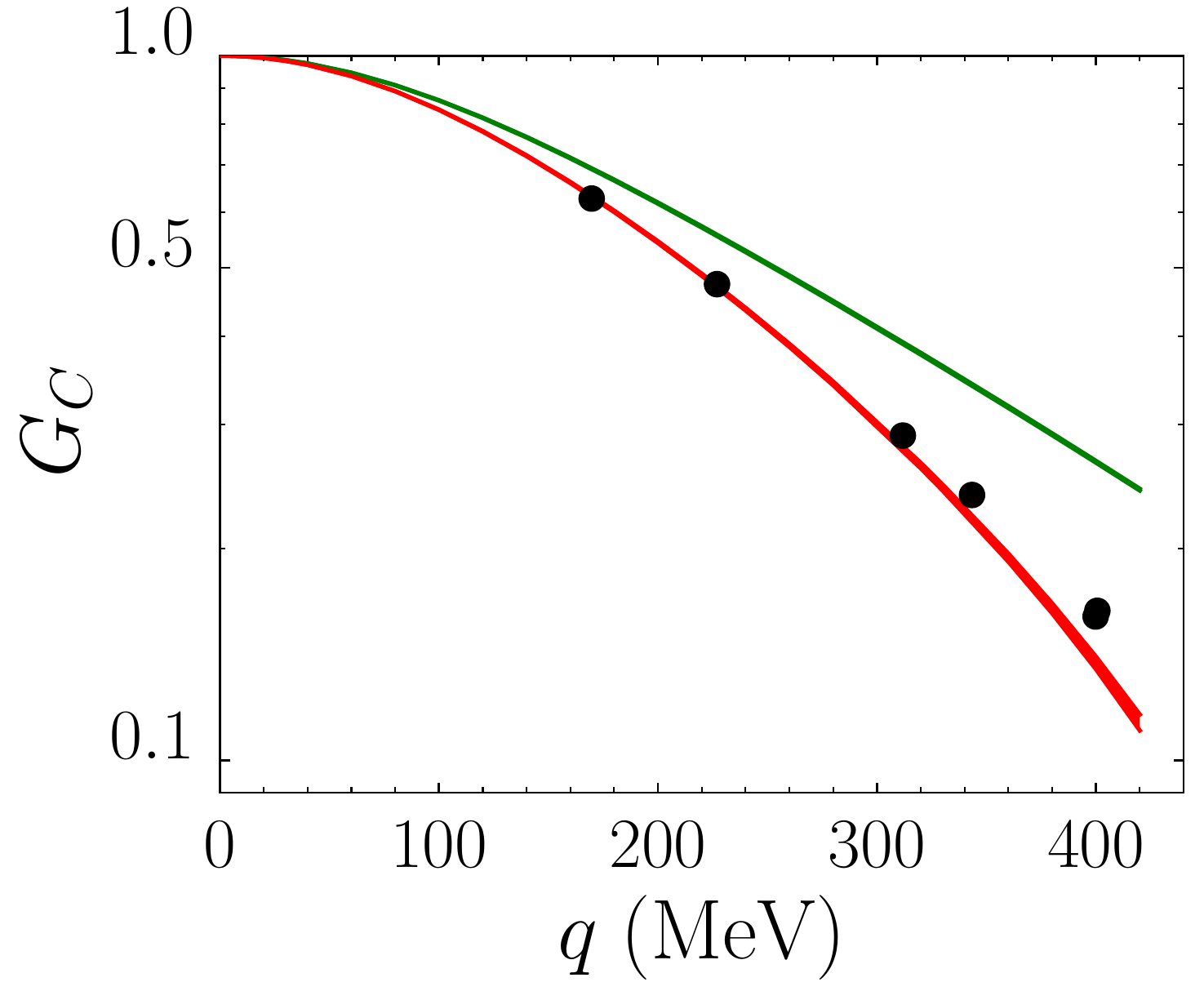}
\includegraphics[scale=0.4]{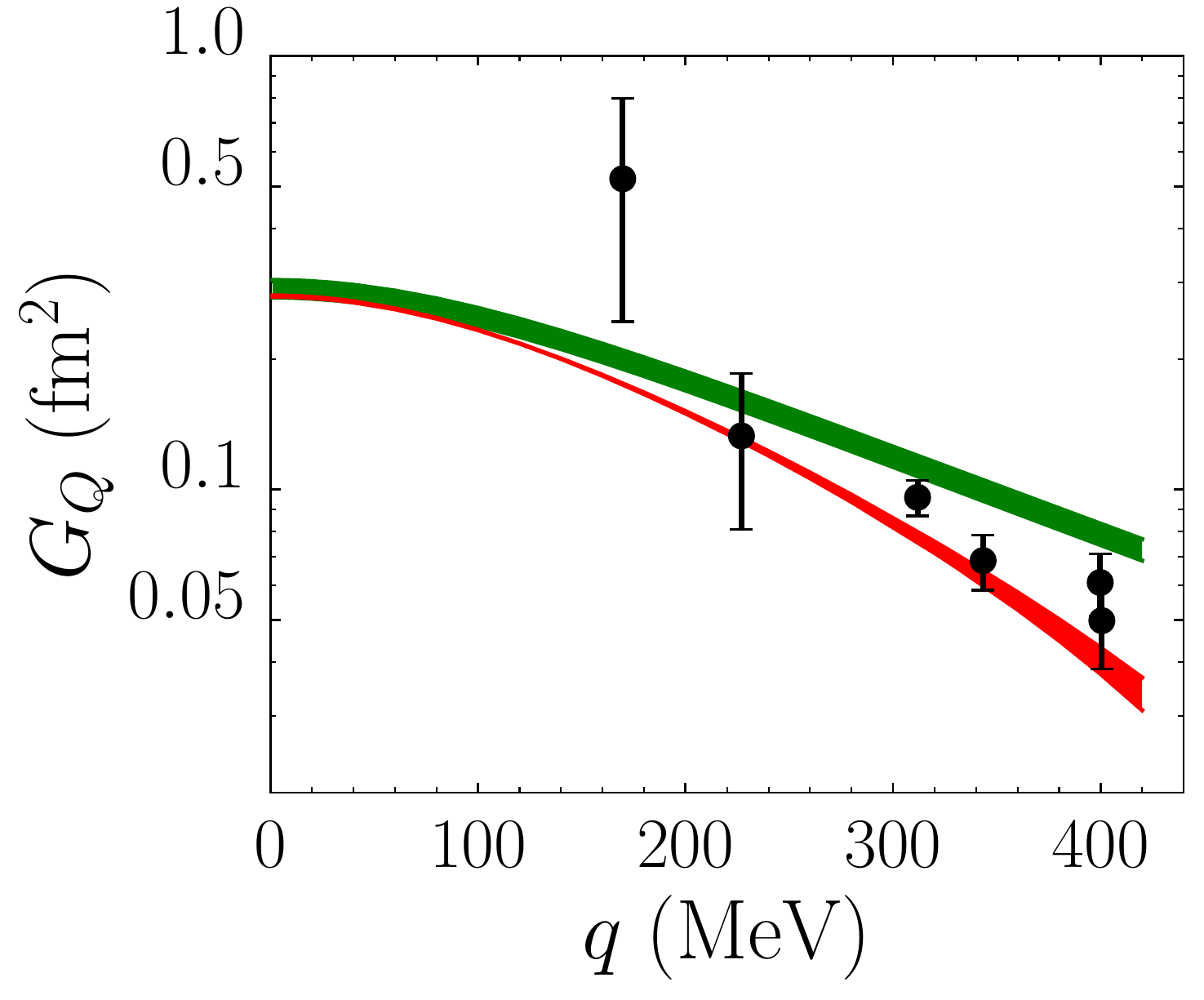}
\includegraphics[scale=0.4]{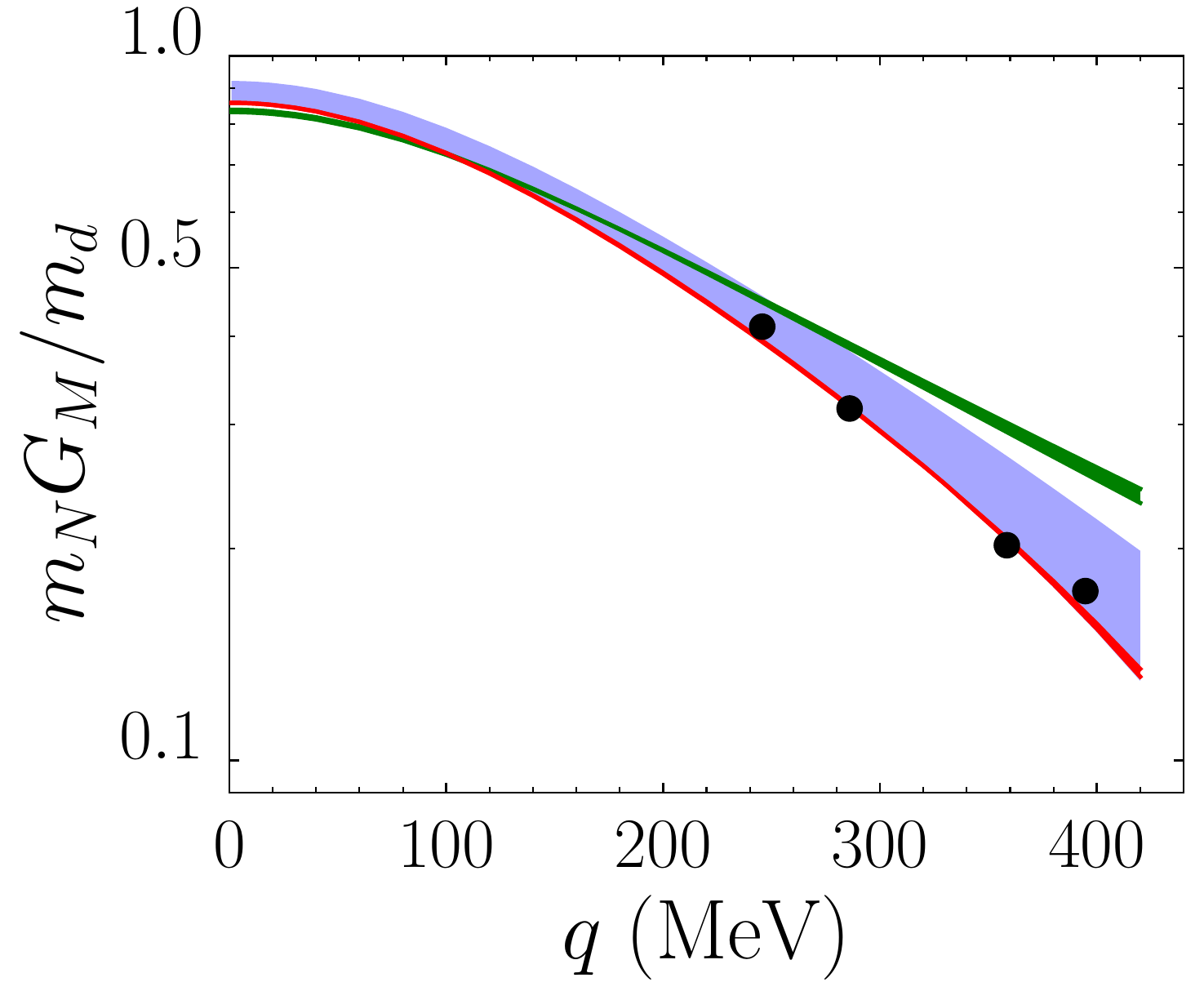}
\caption{The deuteron charge, electric quadrupole, and magnetic form factors as a function of $q$. The green (red) bands represent their LO ({\NNLO}) using two cutoff values $\Lambda = 600$ and $2400$ MeV. For the {\NNLO} $G_M$, the contact current operator ~\eqref{eqn:NNNLOcurrent_ct_NFFCh} is accounted for. The light blue band represents the {\NNLO} $G_M$ before renormalization. The solid circles and error bars represent the experimental data~\cite{JLABt20:2000qyq, Simon:1981br}. However, for $G_C$ and $G_M$ the scales are too large for the error bars to be visible.}
\label{fig:GCGQGMvsq}
\end{figure}

We now turn to the LO and {\NNLO} form factors as a function of $q$. They are plotted with $\Lambda = 600$ and $2400$ MeV, shown as bands in Fig.~\ref{fig:GCGQGMvsq}. For $G_M$, we use different colors to distinguish and compare before and after renormalization. With the counterterm, the cutoff uncertainty is significantly reduced. The experimental values of $G_C$ and $G_Q$ are taken from Ref.~\cite{JLABt20:2000qyq} and $G_M$ from Ref.~\cite{Simon:1981br}. The order-by-order convergence is clearly demonstrated, and agreement with the experimental data can be achieved at {\NNLO} for $q \lesssim 320$ MeV.

For the charge and electric quadrupole form factors whose {\NNLO} corrections need no counterterms, we can study how contribution from each of the four categories compares with others. In Tables~\ref{tab:GC} and~\ref{tab:GQ}, the LO values and the breakdown of {\NNLO} corrections are shown and $\Lambda = 800$ MeV is used. Although all nominally considered to be {\NNLO}, corrections coming from different mechanisms can scale with different combinations of soft mass parameters: $q$, $m_\pi$, and $\gamma_d$. Indeed, the four contributions vary rather considerably in size from one to another. The nucleon-structure effect is driven by the pion dynamics, so it scales with $m_\pi$ rather than $\gamma_d$ and dominates in both cases of $G_C$ and $G_Q$. One expects the potential correction to be larger than the relativistic one, since chiral nuclear forces likely work below a momentum scale that is smaller than the nucleon mass: $\mhi < m_N$. This is indeed the case for $G_C$ and $G_Q$. The boost correction is suppressed by $\eta \simeq q^2/16m_N^2$, thus is even smaller than the relativistic correction in both cases. 

\begin{table}
\tabcolsep 3mm 
\caption{LO values of and various {\NNLO} corrections to $G_C$ ($\Lambda = 800$ MeV).}
\vspace{2mm}
\centering
\begin{tabular}{ccccccc}
\hline
\hline
        $q$ & LO & Potential & Relativistic & Structure & Boost \\
\hline
      0.5$m_\pi$ & 0.929  & $-0.746\times10^{-3}$ & $-0.435\times10^{-3}$ & $-1.14\times10^{-2}$ & $0.222\times10^{-4}$  \\
      $m_\pi$ & 0.768 & $-2.52\times10^{-3}$ & $-1.42\times10^{-3}$ & $-3.78\times10^{-2}$ & $2.39\times10^{-4}$   \\ 
      1.5$m_\pi$ & 0.596 & $-4.66\times10^{-3}$ & $-2.42\times10^{-3}$ & $-6.60\times10^{-2}$ & $7.43\times10^{-4}$   \\ 
      2$m_\pi$ & 0.448 & $-6.87\times10^{-3}$ & $-3.13\times10^{-3}$ & $-8.81\times10^{-2}$ & $14.4\times10^{-4}$   \\
\hline
\hline
\end{tabular}
    \label{tab:GC}
\end{table}

\begin{table}
\tabcolsep 3mm 
\caption{LO values of and various {\NNLO} corrections to $G_Q$ (in fm$^2$ and $\Lambda = 800$ MeV).}
\vspace{2mm}
\centering
\begin{tabular}{ccccccc}
\hline
\hline
        $q$ & LO & Potential & Relativistic & Structure & Boost \\
\hline
      0.5$m_\pi$ & 0.271  & $-9.18\times10^{-3}$ & $-1.97\times10^{-3}$ & $-0.333\times10^{-2}$ & $0.065\times10^{-4}$  \\
      $m_\pi$ & 0.224 & $-8.00\times10^{-3}$ & $-2.13\times10^{-3}$ & $-1.10\times10^{-2}$ & $0.694\times10^{-4}$   \\  
      1.5$m_\pi$ & 0.174 & $-6.71\times10^{-3}$ & $-2.27\times10^{-3}$ & $-1.92\times10^{-2}$ & $2.13\times10^{-4}$   \\  
      2$m_\pi$ & 0.132 & $-5.57\times10^{-3}$ & $-2.34\times10^{-3}$ & $-2.60\times10^{-2}$ & $4.02\times10^{-4}$   \\
\hline
\hline
\end{tabular}
    \label{tab:GQ}
\end{table}

To disentangle the impacts of multiple soft scales, we remove the explicit dependence on $q$ from the stage by studying the deuteron charge radius and electric quadrupole moment. The potential correction is controlled by $\gamma_d^2/\mhi^2$ because the $m_\pi$-dependence of the nuclear forces has been absorbed into chiral LECs through renormalization, such as $g_A$ and $f_\pi$. The relativistic correction is expected to be suppressed by $\gamma_d^2/m_N^2$. In Tables~\ref{tab:rd2} and ~\ref{tab:Qd}, the {\NNLO} contributions to $r_d^2$ and $Q_d$ by various mechanisms are tabulated, together with the LO values for comparison. The boost correction vanishes for both $r_d^2$ and $Q_d$ and the structure correction vanishes for $Q_d$. 

\begin{table}
\tabcolsep 3mm 
\caption{The LO value of and the {\NNLO} corrections to $r_{d}^{2}$ (in fm$^{2}$ and $\Lambda = 800$ MeV).}
\vspace{2mm}
\centering
\begin{tabular}{cccc}
\hline
\hline
        LO & Potential & Relativistic & Structure  \\
\hline
      3.79 & $3.93\times10^{-2}$ & $2.31\times10^{-2}$ & 0.604 \\
\hline
\hline
\end{tabular}
    \label{tab:rd2}
\end{table}

\begin{table}
\tabcolsep 3mm
\caption{The LO value of and the {\NNLO} corrections to $Q_d$ (in fm$^{2}$ and $\Lambda = 800$ MeV).}
\vspace{2mm}
\centering
\begin{tabular}{ccc}
\hline
\hline
     LO & Potential & Relativistic \\
\hline
     0.291 & $-9.70\times10^{-3}$ & $-1.89\times10^{-3}$ \\
\hline
\hline
\end{tabular}
    \label{tab:Qd}
\end{table}

The potential corrections are expected to be around $(\gamma_d/\mhi)^2$. For $r_d^2$, this translates into $\mhi \approx 500$ MeV, more or less in agreement with the expectation from the literature~\cite{Epelbaum:2019kcf, Hammer:2019poc}. But $Q_d$ points to a much smaller $\mhi \approx 260$ MeV. Because $Q_d$ and the mixing angle $\varepsilon_1$ are closely correlated, the slow convergence of $Q_d$ probably reflects the slow convergence of the chiral forces for $\varepsilon_1$ in comparison with, say, the $\cs{3}{1}$ phase shift~\cite{Long:2011xw}. The relativistic corrections are expected to be $(\gamma_d/m_N)^2 \simeq 0.25\%$. The actual values of the relativistic corrections turn out to be larger than this naive expectation, $0.61\%$ for $r_d^2$ and $0.65\%$ for $Q_d$, although still smaller than the potential corrections in magnitude. 

\section{Summary\label{sec:summary}}

We have calculated the deuteron charge, electric quadrupole, and magnetic form factors up to {\NNLO} in nuclear ChEFT, treating all subleading corrections in strict perturbation theory. The cutoff variations of these form factors were studied with the intent to examine the NDA counting assigned to the two-body contact charge and current operators. This is the first of a line of studies to investigate the difference of NDA and the power counting proposed by Ref.~\cite{PavonValderrama:2014zeq} in light of perturbative calculations.

Up to the order worked out, we found that the contact current operator~\eqref{eqn:NNNLOcurrent_ct_NFFCh} contributing to the deuteron magnetic form factor needs to be promoted to {\NNLO}, as opposed to N$^3$LO in NDA. This agrees with Ref.~\cite{PavonValderrama:2014zeq} in that this particular current operator is enhanced relative to NDA. Combining $\vec{J}_\text{ct}$ with {\NNLO} long-range contributions, we find that the magnetic form factor is properly renormalized.

For the deuteron charge and quadrupole form factors, satisfactory cutoff independence was observed and there is no indication for any necessity to modify NDA counting. In these two cases, the {\NNLO} corrections were categorized according to their source and were analyzed against the scaling expected of them.

\acknowledgments

We thank Bira van Kolck for encouragement. This work was supported by the National Natural Science Foundation of China (NSFC) under Grant No. 11735003 (BL) and the Fundamental Research Funds for the Central Universities (LS).

\appendix
\section{Integrals for evaluating \texorpdfstring{$G_C$}{GC}, \texorpdfstring{$G_Q$}{GQ}, and \texorpdfstring{$G_M$}{GM}}\label{sec:analyticexpressions}

The integral formulas for the matrix elements of various charge and current operators between the deuteron wave functions are spelled out below, in addition to Eqs.~\eqref{eqn:LOGC_integral} and \eqref{eqn:LOGQ_integral}. Besides the expression of the operators and the notation of the form factors, one can also find in Sec.~\ref{sec:formalism} the definitions for variables such as $p'$, $x$, $y$, and $z$.

\begin{align}
\begin{split}
     G_{M}
     \left(Q^2| \vec{J}\,^{(0)} \right) ={}& \frac{m_d}{32\,q\,m_{N}} \int_{0}^{\infty}dp \int_{-1}^{1}dz\, \frac{p}{p^{\prime 3}} \bigg{\{}16 (1+\kappa_s) p^{\prime 2} q u(p')u(p)\\
&- 8\sqrt{2}p^{\prime 2} \left[3pz \left(z^{2} - 1\right) + 2 (1+\kappa_s) q P_2(z) \right] u(p')w(p)\\
&+ 4\sqrt{2}\left[ 6p^{2}p'y\left(1 - z^{2}\right) + 2 (1+\kappa_s) p^{\prime 2} q P_2(y)  \right] w(p')u(p)\\
&- \Big{\{}12p \left(z^2 -1\right) \left[p'y(2p + 3qz) - 4p^{\prime 2}z\right] \\
&+ 4 (1+\kappa_s) p'q \left[6pyz + \left(3qy - 2p'\right) P_2(z) \right] \Big{\}} w(p')w(p)\bigg{\}} \, .
\end{split} 
\end{align}

\begin{equation}
       G_{C}
       \left(Q^2| \rho^{(2)}_\text{rel}\right) = - \frac{1 + 2\kappa_{s}}{24\, m_{N}^{2}}\left[\mathcal {A}
    + 3 Q^2 G_{C}(Q^2| \rho^{(0)}) \right] \, ,    
\end{equation}
where
\begin{equation}
\begin{split}
     \mathcal {A} \equiv{}& \frac{9\,q}{2} \int_{0}^{\infty}dp\int_{-1}^{1}dz\, \frac{p^2}{p^{\prime 3}} \left(z^2 - 1\right) \left[p'y(qz + p) - 
    p^{\prime  2} z\right] w(p') w(p) \, .
\end{split}
\end{equation}
\begin{equation}
G_{Q}
  \left(Q^2| \rho^{(2)}_\text{rel}\right) = - \frac{1 + 2\kappa_{s}}{4 m_{N}^{2}} \left[ \frac{\mathcal {B}}{q^2} + \frac{Q^{2}}{2}  G_{Q}(Q^2| \rho^{(0)}) \right] \, ,
\end{equation}
 where
\begin{equation}
\begin{split}
  \mathcal {B} \equiv{}& \frac{9\, q}{8} \int_{0}^{\infty} dp \int_{-1}^{1} dz\, \frac{p^2}{p^{\prime  3}}
    \left(1 - z^2\right) \Big{[}2\sqrt{2}\, p^{\prime 2}\, z\, u(p')w(p) \\
    &- 2\sqrt{2} pp'y w(p')u(p) - p'qyz\, w(p')w(p)\Big{]} \, .
\end{split}
\end{equation}
\begin{equation} 
\begin{split}
  G_{M}\left(Q^2| \vec{J}_\text{rel}^{\,(2)} \right) ={}& \frac{\sqrt{2}\, m_{d}}{8\, q} \int_{0}^{\infty} dp \int_{-1}^{1} dz\, \frac{p}{p'} \Big{[}8\,\mathcal{C}\, u(p') u(p) + \sqrt{8} \left(3\,\mathcal{E} - \mathcal{C}\right) u(p') w(p)     \\
&+ \sqrt{8} \left(3\,\mathcal{D} - \mathcal{C}\right) w(p') u(p) + \left(\mathcal{C} - 3\,\mathcal{D} -3\,\mathcal{E} + 9\mathcal{F}\right) w(p') w(p) \Big{]}   \, ,
\end{split}
\end{equation}
where 
\begin{align} 
    \begin{split}
  \mathcal{C} \equiv{}& -\frac{q}{32\sqrt{2}\, m_{N}^{3}} \left[4p^{2} \sin^{2}\theta + (1 + \kappa_s) \left(10p^{2} + 3q^2 + 8pq\cos\theta + 6p^{2} \cos 2\theta\right) \right]    \, ,  \\
  \mathcal{D} \equiv{}& -\frac{1}{64\sqrt{2}\, m_{N}^{3}} \Big{\{}p\sin\theta \left[\left(16p^{2} + q^2\right) \sin 2\phi + 8pq \sin(2\phi+\theta)\right]  \\
    &+ 2 (1 + \kappa_s) q \big{[}\cos^{2}\phi \left(10p^{2} + 3q^2 + 8pq\cos\theta + 6p^{2}\cos 2\theta \right)   \\
    &+ 4pq \cos\phi \sin\phi \sin\theta + 4p^{2} \sin^{2}\phi \sin^{2}\theta \big{]} \Big{\}}    \, ,   \\
    \mathcal{E} \equiv{}& -\frac{1}{64\sqrt{2}\, m_{N}^{3}} \Big{\{}2p\sin^{2}\theta \left[\left(16p^{2} + q^2\right) \cos\theta + 4pq\right]    \\
    &- (1 + \kappa_s) q \left[7pq \cos\theta + \left(20p^{2} + 6q^2\right) \cos 2\theta + 9pq \cos 3\theta + 8p^{2} + 4p^{2} \cos 4\theta \right] \Big{\}}   \, ,    \\
     \mathcal{F} \equiv{}& \frac{1}{64\sqrt{2}\, m_{N}^{3}} \Big{\{}p\sin\theta \left[\left(16p^{2} + q^2\right) \sin(2\phi - 2\theta) + 8pq \sin(2\phi - \theta)\right]\\
    &+ (1+\kappa_s) q \big{[}4p^{2} + 4p^{2} \cos 2\phi + 6pq \cos(2\phi - 3\theta) + 4p^{2} \cos(2\phi - 4\theta)\\
    &+ \left(8p^{2} + 3q^2\right) \cos(2\phi - 2\theta) +  2pq \cos(2\phi - \theta)\\
    &+ 5pq \cos\theta + \left(12p^{2} + 3q^2\right) \cos 2\theta + 3pq\cos 3\theta \big{]}\Big{\}}    \, ,  
\end{split}
\end{align}
with
\begin{equation} 
\begin{split}
\phi \equiv \arccos y\, , \quad
\theta \equiv \arccos z \, .
\end{split}
\end{equation}
\begin{equation}
\begin{split}
    G_{M}
     \left(Q^2 | \vec{J}_\text{str}^{\,(2)}\right) ={}& -\frac{m_d}{192\,q\,m_{N}} \int_{0}^{\infty}dp \int_{-1}^{1}dz\, \frac{p}{p^{\prime 3}} \Big{\{}16 \left\langle r_{MS}^2 \right\rangle (1+\kappa_s) p^{\prime 2} q^3 u(p')u(p)\\
&- 8\sqrt{2}p^{\prime 2} \left[3 \left\langle r_{ES}^2 \right\rangle pq^2z \left(z^{2} - 1\right) + 2 \left\langle r_{MS}^2 \right\rangle (1+\kappa_s) q^3 P_2(z) \right] u(p')w(p)\\
&+ 4\sqrt{2}\left[ 6 \left\langle r_{ES}^2 \right\rangle p^{2}p'q^2y \left(1 - z^{2}\right) + 2 \left\langle r_{MS}^2 \right\rangle (1+\kappa_s)p^{\prime 2} q^3 P_2(y)  \right] w(p')u(p)\\
&- \big{\{}12 \left\langle r_{ES}^2 \right\rangle pq^2 \left(z^2 -1\right) \left[p'y(2p + 3qz) - 4p^{\prime 2}z\right] \\
&+ 4 \left\langle r_{MS}^2 \right\rangle (1+\kappa_s)p'q^3 \left[6pyz + \left(3qy - 2p'\right) P_2(z) \right] \big{\}} w(p')w(p)\Big{\}}    \, .
\end{split}   
\end{equation}

\begin{align} 
   G_M\left(Q^2 | \vec{J}_\text{ct}\right) = \pi\, m_{d}\, L_2 \left(4 I_1^2 - 2 \sqrt{2} I_1 I_2 
 - 4 I_2^2\right)  \, ,
\label{eq:contact_current_GM}
\end{align}
where 
\begin{align} 
\begin{split}
 I_1 \equiv{}& \int_{0}^{\infty}dp \int_{-1}^{1}dz\, p u(p)  \, , \\
 I_2 \equiv{}& \int_{0}^{\infty}dp \int_{-1}^{1}dz\, p P_2(z) w(p)  \, .
\end{split}
\end{align}

\bibliography{Refs_GCGQGM.bib}

\end{document}